\documentclass[a4paper,11pt]{article}
\pdfoutput=1 

\usepackage{jinstpub} 

\usepackage{soul}
\usepackage{color}
\usepackage[dvipsnames]{xcolor}
\usepackage{multicol} 
\usepackage{array} 
\usepackage{wasysym}
\usepackage{hyperref}
\usepackage{url}
\usepackage{color}\definecolor{darkblue}{rgb}{0,0,1}
\hypersetup{
    unicode=false,          
    pdftoolbar=true,        
    pdfmenubar=true,        
    pdffitwindow=true,      
    pdftitle={My title},    
    pdfauthor={Author},     
    pdfsubject={Subject},   
    pdfnewwindow=true,      
    pdfkeywords={keywords}, 
    colorlinks=true,       
    linkcolor=darkblue,          
    citecolor=darkblue,        
    filecolor=magenta,      
    urlcolor=cyan           
}

\usepackage{eurosym}
\usepackage{pifont}
\usepackage{subfigure}
\usepackage[bottom]{footmisc}
\usepackage{gensymb}
\usepackage{url}
\usepackage{multirow}
\usepackage{comment}
\usepackage{xspace}
\usepackage{siunitx}
\usepackage{graphicx}
\graphicspath{ {Figures/} }

\usepackage{lineno}

\usepackage{tablefootnote}
\usepackage{ragged2e}
\usepackage{booktabs}
\usepackage{mathrsfs}

\newcommand{\oxeq}{[O$_2$]$_{\mbox{\small{eq}}}$}
\newcommand{\Wb}{\ensuremath{\mathscr{W}_{Birk}}\xspace}
\newcommand{\Geff}{\ensuremath{\mathscr{G}_{eff}}\xspace}
\newcommand{\GLEM}{\ensuremath{\mathscr{G}_{LEM}}\xspace}
\newcommand{\trans}{\ensuremath{\cal T}\xspace}

\newcommand{\Eextr}{\ensuremath{E_{\mbox{\tiny{extr}}}^{\mbox{\tiny{liq}}}}\xspace}
\newcommand{\Eind}{\ensuremath{E_{\mbox{\tiny{ind}}}}\xspace}
\newcommand{\Eamp}{\ensuremath{E_{\mbox{\tiny{amp}}}}\xspace}
\newcommand{\Ed}[1]{\ensuremath{E_{\mbox{\tiny{#1}}}}\xspace}

\newcommand{\pilot}{\ensuremath{3\times1\times1\mbox{ m}^3\mbox{ demonstrator}}\xspace}
\newcommand{\three}{\ensuremath{3\times1\times1\mbox{ m}^3}\xspace}

\newcommand{\fifty}{\SI[product-units=power]{50x50}{\cm}\xspace}
\newcommand{\figref}[1]{\figurename~\ref{#1}}
\newcommand{\secref}[1]{Section~\ref{#1}}

\newcommand{\tabref}[1]{Table~\ref{#1}}

\title{Performance study of a \three dual phase liquid Argon Time Projection Chamber exposed to cosmic rays}

\author[a]{B.\,Aimard}
\author[q]{L.\,Aizawa}
\author[w]{C.\,Alt}
\author[b]{J.\,Asaadi}
\author[e]{M.\,Auger}
\author[n]{V.\,Aushev}
\author[u]{D.\,Autiero}
\author[g]{A.\,Balaceanu}
\author[a]{G.\,Balik}
\author[u]{L.\,Balleyguier}
% ...

\affiliation[a]{LAPP, Universit\'e Savoie Mont Blanc, CNRS/IN2P3, Annecy, France}
\affiliation[b]{University of Texas Arlington, Arlington, USA}
% ...
\affiliation[e]{University of Bern, Albert Einstein Center for Fundamental Physics, Laboratory for High Energy Physics (LHEP), Bern, Switzerland}
% ...
\affiliation[g]{Horia Hulubei National Institute for R\&D in Physics and Nuclear Engineering - IFIN-HH, Bucharest - Magurele, Romania}
% ...
\affiliation[n]{Kyiv National University, Kyiv, Ukraine}
% ...
\affiliation[q]{Iwate University, Department of Electrical Engineering and Computer Science, Morioka, Iwate, Japan}
% ...
\affiliation[u]{Institut de Physique des 2 Infinis (IP2I), CNRS/IN2P3, Universit\'e de Lyon, Universit\'e Claude Bernard Lyon 1, Villeurbanne, France}
% ...
\affiliation[w]{ETH Zurich, Institute for Particle Physics, Zurich, Switzerland}

\emailAdd{mario.campanelli@cern.ch}

\abstract{We report the results of the analyses of the cosmic ray data collected with a 4 tonne (3$\times$1$\times$1~m$^3$) active mass (volume) Liquid Argon Time-Projection Chamber (TPC) operated in a dual-phase mode. We present a detailed study of the TPC's response, its main detector parameters and performance.  The results are important for the understanding and further developments of the dual-phase technology, thanks to the verification of key aspects, such as the extraction of electrons from liquid to gas and their amplification through the entire one square metre readout plain, gain stability, purity and charge sharing between readout views.}


\begin{document}
\maketitle
\flushbottom

\section{Introduction} 
\label{sec_intro}

Dual phase (DP) Liquid Argon TPCs (LAr TPC) have been under development for more than a decade~\cite{Badertscher:2008rf,Badertscher:2010fi,Badertscher:2013wm,devis-thesis,filippo-thesis}. 
They combine the properties of the liquid argon as the tracking medium with the charge amplification in the gaseous argon phase above the liquid at the top of the detector, after extracting drift electrons from the liquid into the gas. This provides a full 3D imaging of charged particle interactions with high resolution and low energy threshold. 
As such, they can be a potent tool to study neutrino interactions and other rare phenomena \cite{Gonzalez-Diaz:2017gxo}. 
DP LAr TPCs at the 10 kt scale, located underground and coupled to powerful neutrino beams offer a rich scientific portfolio, that includes the study of leptonic Charge-Parity (CP) violation and the determination of the neutrino mass ordering. Thanks to the VUV scintillation of the liquid argon, they can operate in self-triggering mode to study atmospheric and astrophysical neutrinos with very high statistics and extend the sensitivity to proton decay. 

The possibility to deploy and operate large scale DP LAr TPCs underground has been studied in detail in the context of LAGUNA-LBNO~\cite{Stahl:1457543} and subsequently in DUNE~\cite{Abi:2018dnh}. Two prototype LAr TPCs (known as Dual Phase and Single Phase ProtoDUNE) with about 300 tonnes in active mass each have been constructed and operated at CERN \cite{DeBonis:1692375} by the DUNE collaboration. In 2017, a 4 tonne (\three) DP LAr TPC was operated, collecting over 300,000 cosmic ray interactions. A first look at this data presented in Ref.~\cite{Aimard:2018yxp} clearly demonstrates the excellent imaging capabilities of the detector and provides a proof of principle of the DP technology at the tonne scale.  Follow up studies performed with the scintillation light in the detector have been published in \cite{aimard:2020qqa}.

A more detailed analysis of the ionisation charge signals produced in the pilot detector is the subject of this paper. Based on the acquired cosmic ray data, we evaluate the impact and the relative importance of various parameters on quantities such as signal-to-noise ratio, hit finding efficiency and the uniformity of the response on a $3\times1$\,m$^2$ area. 
The \pilot is the first large scale prototype of a DP LAr TPC and a reduced scale replica of upcoming larger detectors. In this context, a detailed study of its response and imaging capabilities is of utmost importance to better gauge the performance and provide more accurate simulations of large scale DP LAr TPCs. 

In \secref{sec_overview}, we briefly present the overview of the detector along with the response to the charge readout system. We also describe the key detector parameters that can have the largest impact on the quality of the data. In \secref{sec_datamc}, we present some details on the Monte Carlo and data samples, and the algorithms used for track reconstruction. Finally, we evaluate  the detector response and performance in \secref{sec_detperf}.

\section{Overview of the \three demonstrator detector}
\label{sec_overview}
\subsection{Experimental setup}
\begin{figure}[h!]
\begin{center}
\includegraphics[width=1\textwidth,viewport=0 0 650 300]{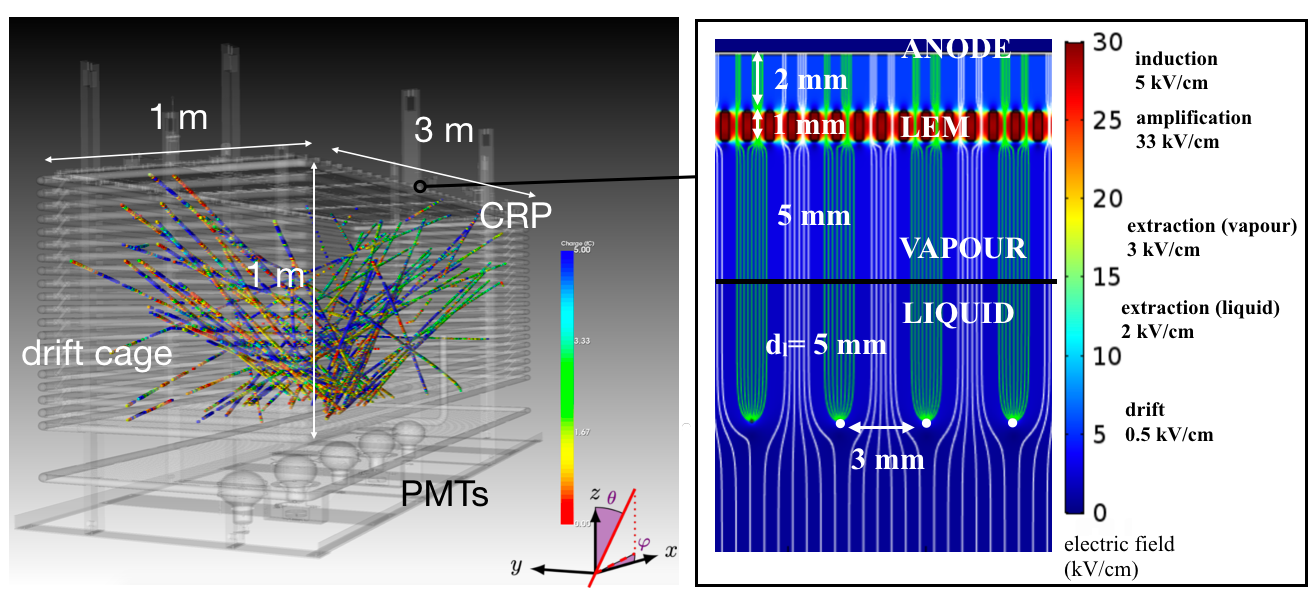}
\caption{Left: The \three geometry with several reconstructed cosmic muon tracks from different events superimposed. Only reconstructed hits from one of the views after the 3D track matching are shown and the colour scale represents the acquired hit charge integral in fC. Right: Nominal electric field settings of the detector.}
\label{fig:311-geometry}
\end{center}
\end{figure}

The \three demonstrator detector is described in detail in Ref.~\cite{Aimard:2018yxp}. It consists of a \three liquid argon active volume, defined by a cathode at the bottom, a field cage, and the readout plane at the top. 
A charge amplification and anode readout stage is positioned in the gas phase a few millimetres above the liquid argon surface. 
\figref{fig:311-geometry} shows several reconstructed cosmic muon track candidates from different events superimposed, along with a schematic of the signal amplification region in the right panel. 


At the passage of an ionising particle inside the liquid, electron-ion pairs are created. The electrons drift towards the surface at a velocity of about \SI{1.6}{\mm/\micro\s}, thanks to the nominal electric field of \SI{500}{\volt/\cm} provided by the cathode and drift cage system. Scintillation photons with wavelengths peaked at \SI{128}{\nm} are also produced and detected by five PMTs placed in line at the bottom of the cryostat. They can provide the trigger and the reference time for the event with an accuracy of a few nanoseconds \cite{aimard:2020qqa}. The produced charge is free to drift in the ultra pure liquid argon up to the surface where it is extracted to the gas phase. Once in the gas, the drifting electrons are multiplied by twelve $50 \times 50$ cm$^2$ Large Electron Multipliers (LEMs) and collected on a two-dimensional and finely segmented anode. The amplification in the LEMs allows to produce signals with amplitudes which are significantly above the level of the ambient noise, resulting in a high quality and good resolution image of the interacting particles. 

The electron extraction, amplification, and collection are performed inside a $3\times1$~ m$^2$ structure called Charge Readout Plane (CRP). The CRP is electrically and mechanically independent from the drift cage and can be remotely adjusted to the liquid level. The electrons are efficiently extracted from the liquid to the vapour by applying an electric field in the liquid above \SI{2}{\kilo\volt/\cm} \cite{Gushchin:1982}. This electric field is provided by a \SI{3}{\mm} pitch extraction grid positioned \SI{5}{mm} below the liquid argon surface. Once amplified inside the LEM holes, the charge is collected on a two-dimensional segmented anode which consists of a set of independent strips that provide the $x$ and $y$ coordinates of an event with a \SI{3.125}{\mm} pitch. The anodes are electrically bridged together so as to provide two orthogonal sets of three and one metre long readout strips called \textit{views}. The anodes are carefully designed in such a way that the amplified charge is equally shared and collected on both views \cite{Cantini:2013yba}. 

\begin{figure}[h!]
\begin{center}
\includegraphics[width=0.9\textwidth,viewport=100 0 1100 370]{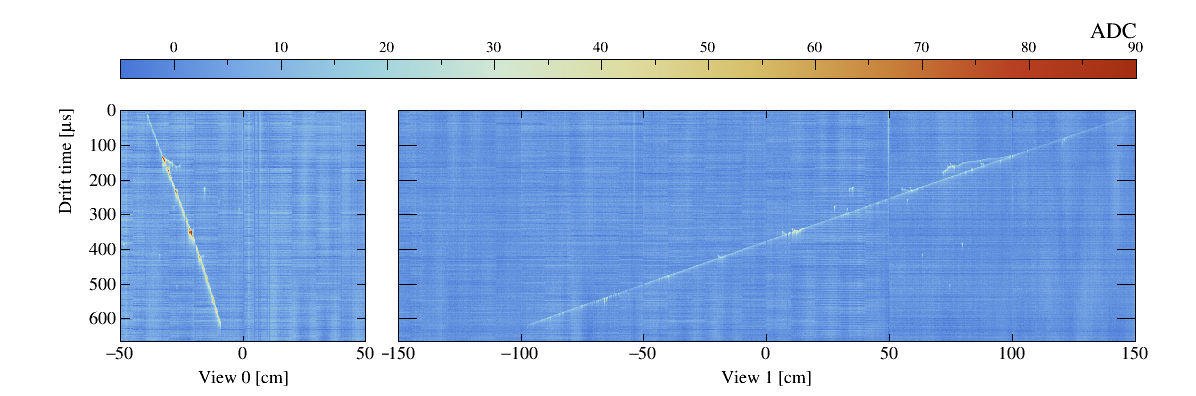}
\caption{2D event display of a cosmic muon track candidate. The event is shown without the application of any noise filtering algorithms.}
\label{fig:311-hadr-event}
\end{center}
\end{figure}

View 0 (view 1) consists of 320 three metre-long (960 one metre-long) strips running parallel to the \textit{x} (\textit{y}) coordinate axis. As an example, in \figref{fig:311-hadr-event} a typical raw cosmic track, recorded with the \pilot is shown.
For each view the charge recorded per channel number is plotted as a function of the drift time, and the colour scale, proportional to the strength of the signal, has been optimised for optimal track visualisation. The readout pitch on the horizontal axis is 3.125~mm and the drift time acquisition window of \SI{667}{\micro\s} is digitally sampled at \SI{2.5}{\mega\Hz} which results in 1667 time samples. At a \SI{500}{\volt/\cm} drift field, this leads to a $\sim$\SI{0.64}{\mm} spatial resolution along the drift coordinate. Each readout view therefore provides a high resolution digitised image of the interaction products with a pixel unit size of $\sim3.125\times0.64$~mm$^2$. A list of relevant parameters of the TPC charge readout section along with properties of the liquid and gas argon is provided in \tabref{tab:lar-prop}.

\begin{table}[!ht]
\renewcommand{\arraystretch}{1.2}
\begin{center}
\begin{tabular}{p{.62\textwidth}p{.18\textwidth}p{.2\textwidth}}
  \toprule
  Property & Value & Unit \\
  \midrule
  \textbf{TPC charge readout} & &\\
   active area   & 3 $\times$ 1 & \SI{}{\m\squared}\\
    N LEMs (anodes) & 12 & \\
    readout strip width &3.125 & \SI{}{\mm}\\
    readout strip length View 0 (View 1) & 3 (1)& \SI{}{\m}\\
    N channels View 0 (View 1) & 320 (960)&\\
    drift length (time) &1 (607)& \SI{}{\m} (\SI{}{\us})\\ 
    drift time digital sample & 400 & \SI{}{\ns}\\
    drift time acquisition window & 667 & \SI{}{\us}\\
    N drift time digital samples & 1667&\\
    N Pixels $N_{channel}^{charge} \times N_{sample}^{drift}$ View 0 (View 1)& 0.5 (1.6) & MPixel\\
    
    digital integral to charge conversion View 0 (View 1) & 55 (66) & ADC$\times$tick/fC\\
    $\tau_1,\tau_2$ for View 0 (View 1) & 6.9, 0.5 (2.2,0.9) &\SI{}{\micro\second}\\
    \midrule
 \textbf{liquid and gas argon properties} & &\\   
  ionisation energy $W_{ion}$ \cite{PhysRevA.9.1438}& 23.6$^{+0.5}_{-0.3}$ & \SI{}{\eV}\\
   electron drift velocity \cite{Li:2015rqa} & 1.648  & \SI{}{\mm/\us}\\
   ion drift velocity ($\Ed{drift}<1$ kV/cm)) \cite{pdg,palestini}& 8 & \SI{}{\mm/\s}\\
   average (most probable) energy loss for MIP & 2.1 (1.7) & \SI{}{\mega\eV/\cm}\\
   \Wb for MIP & 71 & \%\\
   $\sigma_{T}$ ($\sigma_{L}$) for 1 m drift \cite{Li:2015rqa}& 1.2 (0.9) & \SI{}{\mm}\\
   $\mathscr{A}$ at 100 ppt (30 ppt) for 1 m drift & 82 (94)&\% \\
    GAr temperature near LEMs during data taking \cite{Aimard:2018yxp} &90$\pm0.5$ & \SI{}{\K}\\ 
    cryostat pressure during data taking \cite{Aimard:2018yxp} &1000$\pm 1.4$ & \SI{}{\milli\bar}\\
     \midrule
    First Townsend coeff. $\alpha$ for \Ed{LEM}= 28 (33) \SI{}{\kilo\volt/\cm}& 20.2 (44.1) & \SI{}{\per\cm}\\
  \bottomrule
        \end{tabular}
     \end{center}
     \caption{\label{tab:lar-prop} Some parameters related to the imaging of the TPC along with a summary of coefficients used for the analysis in this paper. Unless otherwise indicated, all values are taken from Ref. \cite{NIST:LAR}, computed for a liquid with temperature of 87.3 K, pressure of 1 atm, and drift electric field of $\Ed{drift}= \SI{500}{\volt/\cm}$. For the first Townsend coefficient we use the value given in \cite{Shuoxing-thesis} for a pressure of 980 mbar and temperature of 87K. }
   \end{table}
   
\subsection{Effective gain and signal-to-noise ratio}
\label{sec:effgain}
An ionising particle depositing an energy $\Delta E_{dep}$ in the liquid, produces an ionisation charge, free to drift in the liquid,  $Q_0=e(\Delta E_{dep}/W_{ion}) \cdot \Wb$ where $W_{ion}$ is the work needed to obtain the ionisation and \Wb denotes the remaining fraction of produced electrons that do not immediately recombine with the ions. The latter depends on the ionisation density and the drift field and is well parametrised by Birk's Law \cite{tagkey1964iii}. The values $W_{ion}$ and \Wb are reported in \tabref{tab:lar-prop}. At a drift electric field \Ed{drift} of \SI{500}{\volt/\cm}, a minimal ionising particle (MIP) produces a charge yield of about \SI{1}{\femto\coulomb/\mm} after recombination.
The charges drift for a duration $t_{drift}$ and electron attachment to impurities in the liquid attenuates the signal by a factor $\mathscr{A}=e^{-t_{drift}/\tau_e}$ where $\tau_e$ is called the electron lifetime. Its value is dependent on the performance of the cryogenic purification and recirculation system. As shown in \cite{Aimard:2018yxp}, a study of the data for this detector suggests a lifetime $\tau_e$ better than 4 ms, or an oxygen equivalent\footnote{$\tau_e[\mbox{ms}]\approx$300/\oxeq [ppt]} impurity level of $\sim$75 ppt \oxeq.

The amount of charge from a MIP collected on an anode strip of view $i$ can therefore be written as:
\begin{equation}\label{eq:dq}
 Q_{i}^{mip} [\mbox{fC}]\approx Q^{mip}_{proj} [\mbox{fC}] \cdot \mathscr{A}(\tau_e) \cdot \Geff \cdot \frac{1}{2},
\end{equation}
where the effect of transverse and longitudinal diffusion ($\sigma_{T}$,$\sigma_{L}$) is neglected for the relatively short 1 metre drift distance (see \tabref{tab:lar-prop}). The space resolution values are below the channel pitch.
The factor 1/2 represents the fact that the charge yield is equally split between both collection views of the anode.  The projected charge on one anode strip, $Q_{proj}^{mip}$, depends on the azimuthal and polar angles of the track crossing the TPC fiducial volume.


\Geff~is the \textit{effective gain} of the dual phase chamber which primarily relies on the multiplication factor of the electrons inside the LEM holes but also depends on the overall electron transparency (${\cal T}\leq1$) of the extraction grid, LEM and anode inside the CRP. The effective gain can be analytically decomposed as:
\begin{eqnarray}\label{eq:gain_func}
\Geff&=&{\cal T}\cdot \GLEM~\label{eq:g_eff}\\
\GLEM&=& \exp\left[{\alpha\cdot x}\right]\\
\alpha&=&A\rho \cdot \exp\left[{-B \rho/\Ed{LEM}}\right],
\end{eqnarray}
where $\alpha$ is the first Townsend ionisation coefficient for an amplification field \Ed{LEM} inside the LEM hole and gas density $\rho$; $x$ denotes the effective amplification length. $A$ and $B$ are parameters which depend on the gas properties and are obtained from numerical calculations \cite{magboltz}, \cite{Shuoxing-thesis}. The value of $\alpha$ along with the nominal pressure and temperature of the gas argon during operation of the detector are reported in \tabref{tab:lar-prop}. 

Stable effective gains of around 20 have been reported inside litre-scale dual phase TPCs \cite{Cantini:2013yba,Cantini:2014xza} by operating the CRP at the electric field settings shown in \figref{fig:311-geometry}. The \pilot is however limited in effective gain due to the high voltage issues on the grid and LEMs reported in \cite{Aimard:2018yxp}. The results discussed in \secref{sec:geff} show that data has been collected at $\Geff=1.9$. 
A MIP crossing the TPC at the polar and azimuthal angles of ($\theta,\phi$)=($90^\circ,45^\circ$) will produce on average $Q_{i}^{MIP}\approx2.1$~fC of charge on the 3~mm strips (without amplification) and neglecting attenuation due to electron lifetime. 
The effective gain drives the signal-to-noise ratio of the charge readout of the TPC ($S/N$) which is defined as the ionisation charge collected on one anode strip of view $i$ divided by the Equivalent Noise Charge (ENC), the RMS of the noise of that channel.

\begin{equation}
    S/N_i=\frac{Q_{i}^{mip} \mbox{[fC]}}{\mbox{ENC [fC]}}\equiv\frac{\mbox{ionisation charge collected on one anode strip}}{\mbox{ENC}}.
\label{eq:s/n}
\end{equation}

 The ENC on a channel is the combination of the intrinsic noise due to the input capacitance of the anode strips and all other sources of coherent or incoherent noise which may be picked up. As shown in \secref{sec:data_summary} an ENC of around 2'000 electrons on view 1 and 2500 electrons on view 0 is reported at the beginning of the TPC operation period, those values drop to around 1'000 electrons for each view after the application of coherent noise removal.  We note that the amount of collected charge per strip and hence the quoted value of the $S/N$ is  dependent on the track topology. For the rest of the paper, we will quote the $S/N$ per view for the specific case of a MIP track crossing the strips at ($\theta,\phi$)=($90^\circ,45^\circ$) before noise filtering algorithms are applied. This provides a conservative lower bound on the $S/N$ since in general, and especially for cosmic particles, tracks are inclined and will thus have a larger polar angle. 

\begin{figure}[h!]
\begin{center}
\includegraphics[width=\textwidth,viewport=0 0 1742 643]{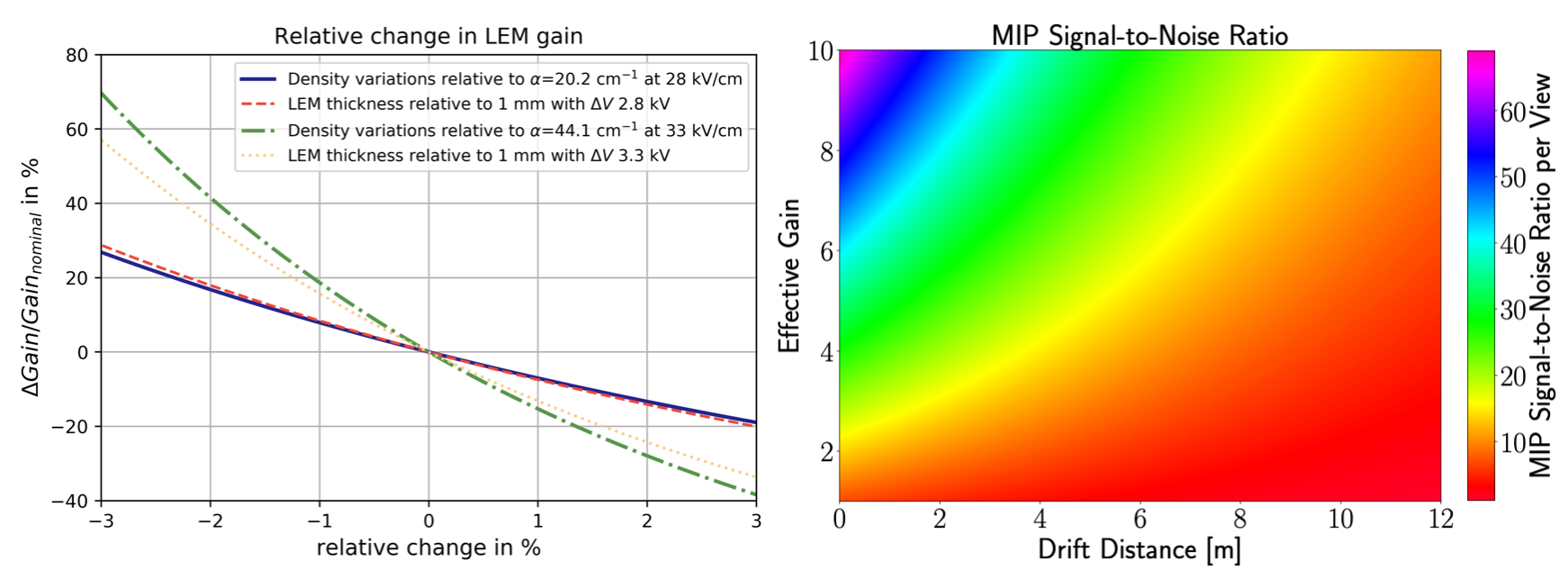}
\caption{Left: relative fluctuation of \GLEM around its nominal value when varying the gas density and LEM thickness from \tabref{tab:lar-prop}. The fluctuation is shown for two different values of \Ed{LEM}. Right: $S/N$ from a MIP interaction collected per view on one anode strip as a function of the effective gain and drift distance. The $S/N$ is computed for a MIP at an angle $\varphi = 45^\circ$, $\theta = 90^\circ$, purity of 75 [O$_2$]$_{eq}$ ppt, drift field of \text{500 V/cm} and noise of ENC = 2000 e$-$.}
\label{fig:gain-var}
\end{center}
\end{figure}

The value of \Geff, as well as its uniformity over the readout surface, therefore directly impacts the imaging quality of the event and is hence a fundamental quantity used to benchmark the detector performance. 
The right side of  \figref{fig:gain-var} illustrates the expected $S/N$ for a MIP crossing a dual phase detector with the operational parameters specified on the figure as a function of \Geff and drift distance.
For the rather modest one metre drift of the \pilot, a $S/N$ of 11.2 can be expected even at an effective gain of 1.9. With similar assumptions for the values of purity and ENC, operating with $\Geff = 10$ would correspond to obtaining $S/N$ = 59.3 for one metre drift and $S/N$ = 11.2 for a TPC with 12 metre drift. Twelve metres is the drift distance of the proposed DUNE 10kt dual phase detector \cite{collaboration2018dune}. The plot illustrates the importance of the effective gain especially if the targets in terms of noise performance or LAr purity are not met.

In order to have a uniform and stable effective gain, the electric field across the LEM, its thickness as well as the density of the argon vapour must be carefully controlled. This is illustrated on the left side of \figref{fig:gain-var}: due to the exponential nature of \Geff, variations in the \fifty LEM thickness or fluctuations in the gas density may have a relatively large impact on its value. In the \pilot, the thickness uniformity of all the \fifty LEM plates was carefully checked and the gas argon density was maintained at a stable value. Both were uniform within typically one percent (see Ref. \cite{Aimard:2018yxp} and \tabref{tab:lar-prop}). The obtained uniformity of the effective gain in the \pilot is discussed in \secref{sec:geff}. In all these studies, diffusion is not accounted for, since over these short drifts its effects are negligible and not measurable. The effects of diffusion, however, should be properly considered in large detectors.

\subsection{Charge readout detector response}\label{subsec:readout_resp}
By design the dual phase TPC, and specifically the \three, allows to place the front-end electronics close to the readout strips, profiting from the cryogenic temperatures while at the same time ensuring its accessibility during detector operation. The  front end electronic boards are described in \cite{Aimard:2018yxp}: they have a DC decoupling stage, high voltage surge protection components and an ASIC chip which contains the CMOS pre-amplifiers and signal shaping. The pre-amplifiers feature a double-slope response with a linear gain for input charges of up to 400 fC and a logarithmic response in the 400-1200 fC range. The double slope feature allows to extend the dynamic range of the detector and acquire events with high ionisation yield. The data presented in this publication is mainly  collected at effective gains of around 2 and therefore we only consider the linear regime of the pre-amplifier response. 

The current produced by the cloud of drifting charge is extracted and multiplied in the LEMs, collected on the anode and pre-amplified. The recorded signal to be digitised $V_{out}(t)$ is a convolution of the induced current $I(t)$ on the anode strip and the response of the pre-amplifier after shaping of the signal $S_{response}(t)$:
\begin{equation}
    V_{out}(t)= \int_{t_0}^{t} I(t')\cdot S_{response} (t-t')~\mbox{d}t'
\end{equation}
with the amplifiers response function:
\begin{equation}
    S_{response}(t)= \frac{\tau_1}{\tau_1 -\tau_2}\cdot \left( e^{-t/\tau_1}-e^{-t/\tau_2} \right),
\label{eq:PreResponse}
\end{equation}
where the values of $\tau_1$ and $\tau_2$ are reported in \tabref{tab:lar-prop}.
The signal, $V_{out}(t)$, after digitisation, called the \textit{waveform}, features a standard shape with a peaking and falling time given by the internal RC-CR shaping performed in the ASIC.
The waveform is characterised by its amplitude, integral and width (defined as the peak full width at half maximum). The integral is proportional to the amount of collected charge.

\begin{figure}[h!]
\begin{center}
\includegraphics[width=0.9\textwidth,height=0.4\textwidth,viewport=0 0 1582 867]{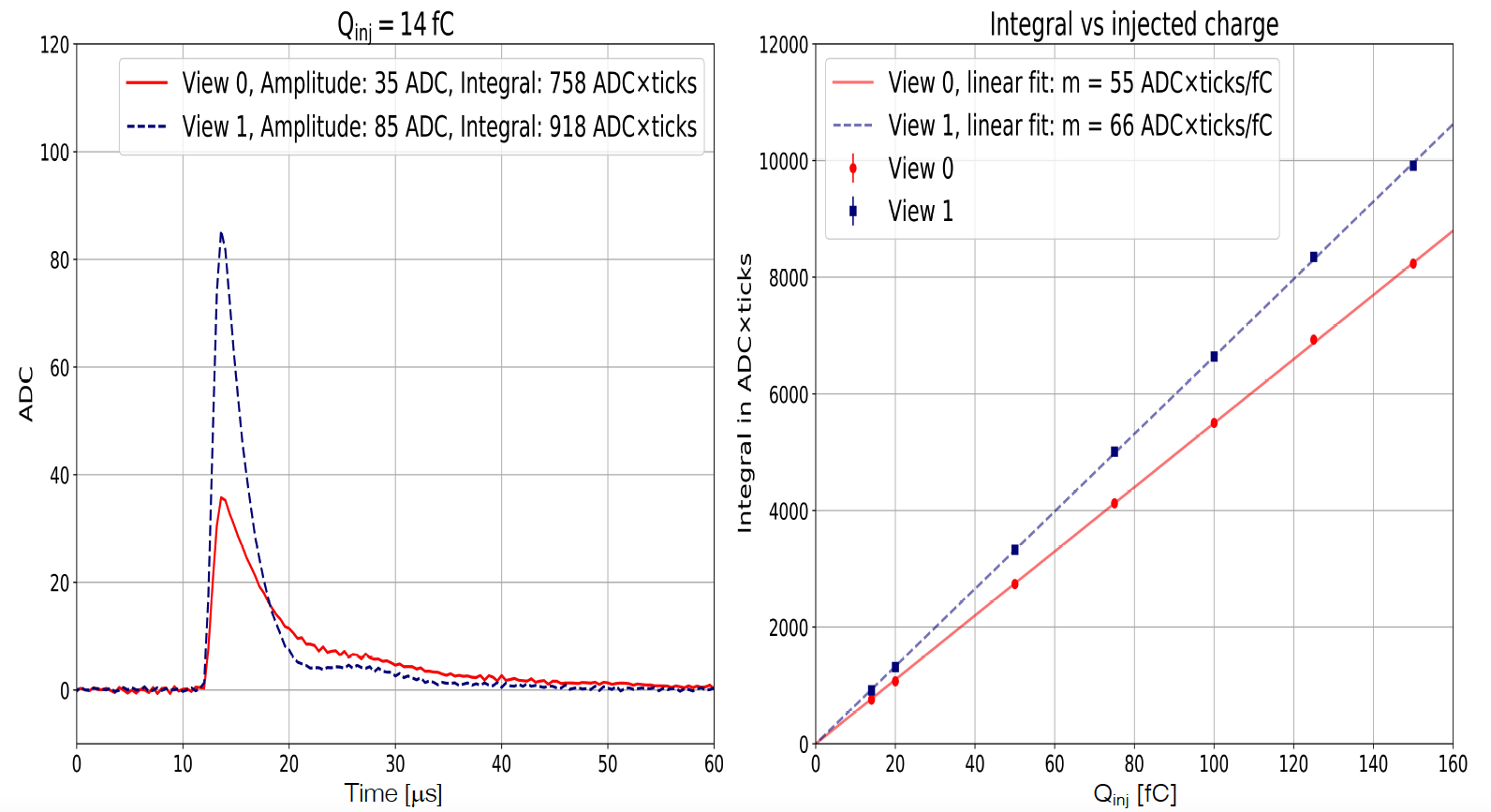}
\caption{Left: response of the pre-amplifiers to an injected pulse of 14 fC on the three metre strips (view 0) and one metre strips (view 1). Right: linearity of the pre-amplifiers response for both views. Ticks refer to time ticks, or time to digital converter (tdc) counts, where each one corresponds to 400 ns.}
\label{fig:pulse-comparison-views}
\end{center}
\end{figure}

Typical waveforms recorded on one channel of view 1 and of view 0 for the same amount of injected charge are shown on the left side of \figref{fig:pulse-comparison-views}.
They are acquired with the system described in \cite{Aimard:2018yxp}, which allows to inject calibrated pulses on groups of 32 strips from each view separately. Since an anode strip has a capacitance to ground of $\sim$160 pF/m \cite{Aimard:2018yxp, Cantini:2013yba} the front end pre-amplifiers connected to the channels of view 0 and view 1 have different capacitive loads. This results in a non-identical pre-amplifier response between both views: for an equal amount of injected charge the waveforms from the channels connected to view 0 have smaller amplitudes, larger widths and a $\sim$20\% lower value of the integral. The linearity of the response in terms of the waveform integral is shown on the right side of \figref{fig:pulse-comparison-views}, where the values of integrals for both views are shown over an extended range of injected charge. The fitted slope for each view is used as a \textit{calibration constant} to convert the digitised integral into fC. The values of both calibration constants are reported in \tabref{tab:lar-prop}. 

\subsection{Charge extraction and electron transparency}\label{subsec:charge_extr}
\begin{figure}[h!]
\begin{center}
\includegraphics[width=0.48\textwidth]{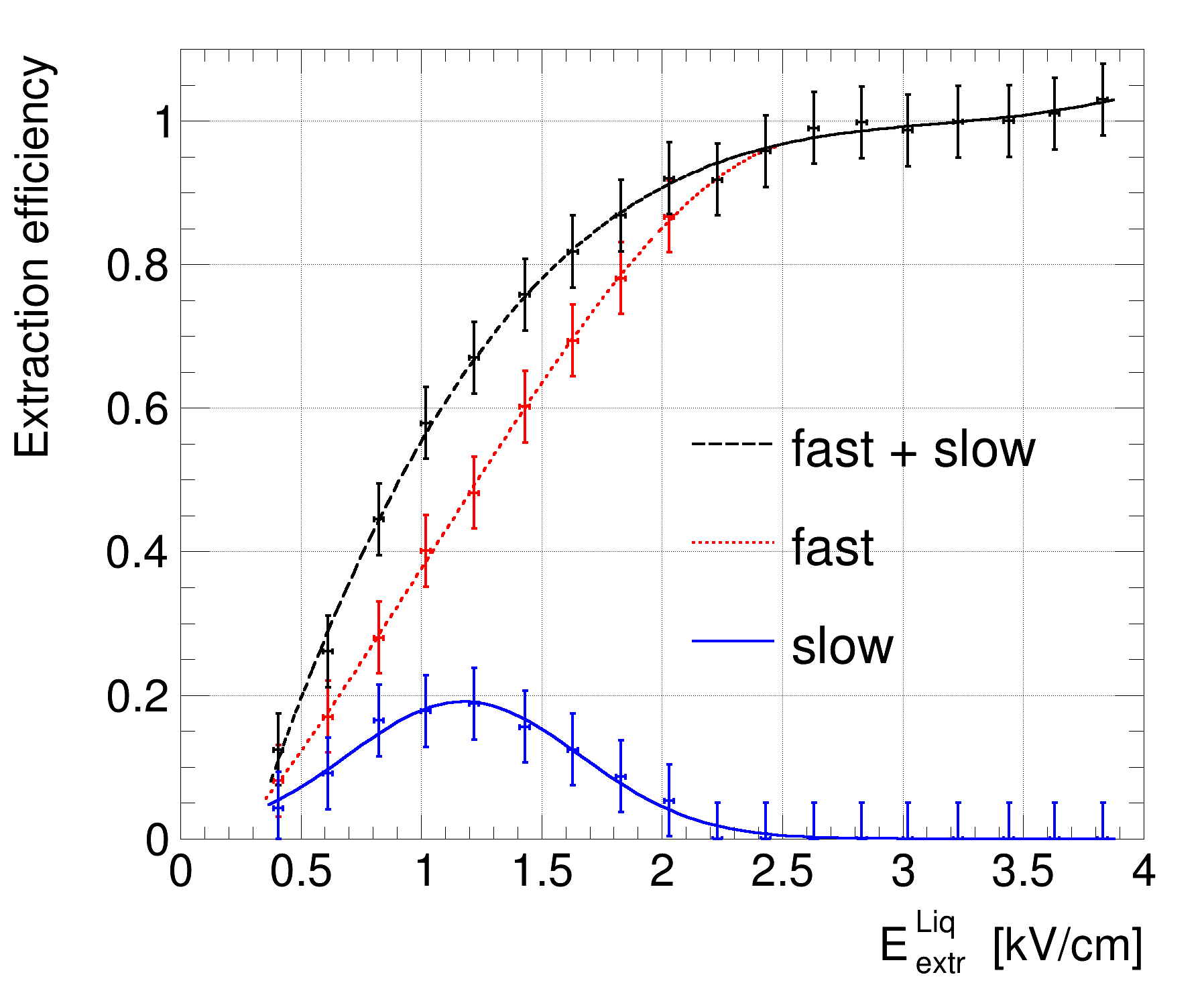}
\includegraphics[width=0.42\textwidth,height=0.4\textwidth]{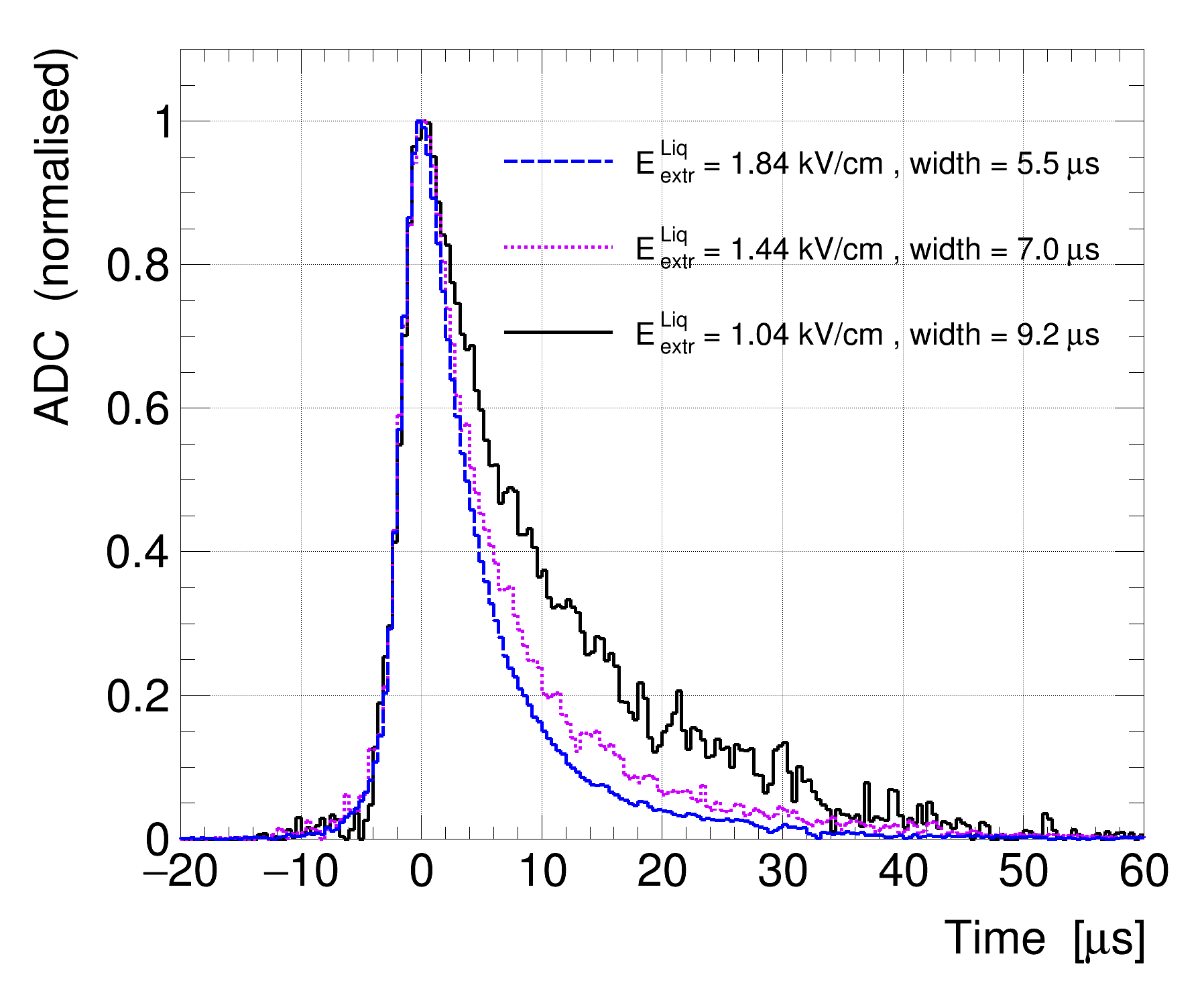}
\caption{Left: efficiency to extract the electrons from the liquid to gas argon phase ($\varepsilon_{extr}^{liq}$) as a function of the applied extraction field in the liquid, $E_{extr}^{liq}$, showing slow and fast components of the signal. The data points are taken from measurements described in \cite{Gushchin:1982}. Right: Waveforms recorded in the \three detector at three different values of the extraction field.}
\label{fig:extr-slow-component}
\end{center}
\end{figure}

For a given LEM amplification field, the electron transparency ${\cal T}\leq1$ is defined as the product of the efficiencies in the extraction and induction regions:
\begin{equation}\label{eq:gain_trans}
{\cal T} = \varepsilon_{trans}\times \varepsilon_{ind},
\end{equation}

where $\varepsilon_{trans}$ denotes the probability for a given charge to be extracted from the liquid and transmitted inside the LEM holes, and $\varepsilon_{ind}$ represents the efficiency of collecting the amplified charges from the top of the LEMs on the anode strips. 
In this paper, we define the transmission efficiency as $\varepsilon_{trans} \equiv \varepsilon_{extr}^{liq} \times \varepsilon_{extr}^{LEM}$ which includes the contributions from the electron transmission from the liquid to the gas phase ($\varepsilon_{extr}^{liq}$) and that of the collection of the electrons entering the LEM from the bottom electrode ($\varepsilon_{extr}^{LEM}$). While the latter is estimated from simulations, the former is taken from previous measurements performed under similar conditions by Gushchin et al. \cite{Gushchin:1982}, shown by the black data points on the left plot of \figref{fig:extr-slow-component}. 

Above 2 kV/cm of extraction field in the liquid (E$_{extr}^{liq}$), most of the charge is extracted within a timescale of less than 100 nanoseconds (``fast extraction'', red markers in \figref{fig:extr-slow-component}), as the electric field decreases a growing fraction of the charge is transmitted via thermionic emission \cite{BORGHESANI1990481}, with characteristic times at the tens of microseconds scale \cite{filippo-thesis}. The impact of the extraction field at which the TPC is operated is noticeable on the shape of the signal as shown in the right panel of \figref{fig:extr-slow-component}, where waveforms from data collected at different extraction fields with similar track topologies, amplification and induction fields are shown. The increase of the width at lower extraction fields due to the slow extraction is clearly visible.

\begin{figure}[h!]
\begin{center}
\includegraphics[width=.8\textwidth]{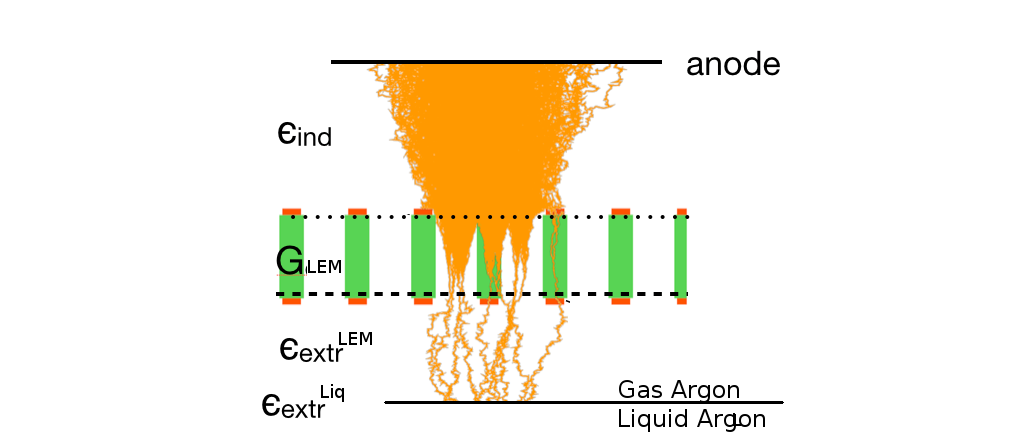}
\caption{Cut view of a 3D Garfield++ simulation of the electron amplification in the LEMs in argon gas. The green and orange rectangles correspond to the electrodes. The LEM holes are illustrated as white spaces. The anode is simplified as a plane of copper at ground potential. The orange lines illustrate the trajectory of the electrons in the gas phase after being extracted from the liquid.}
\label{fig:garfield_transparency}
\end{center}
\end{figure}

It is therefore important not only to operate the TPC at the maximal possible effective gain but also to understand the interplay between the electric fields and estimate the best possible settings in terms of detector transparency. The optimal value of the transparency can be estimated from simulations of electron amplification and transport in the vapour phase. 
In \figref{fig:garfield_transparency}, we show an illustration of a simulated electron avalanche performed with the Garfield++ software package \cite{garfield}. The electric field maps are imported from ANSYS \cite{ansys}, and the gas properties (argon vapour at 90 K) are computed using MAGBOLTZ \cite{magboltz}. 
The 3D simulation is performed over a 3$\times$3 mm$^2$ area, corresponding to one readout pixel. The charge avalanche in the LEM holes is clearly visible in the figure as well as the small fraction of electrons which are collected either on the bottom or top electrodes of the LEM. 

\begin{figure}[h!]
\begin{center}
\includegraphics[width=\textwidth]{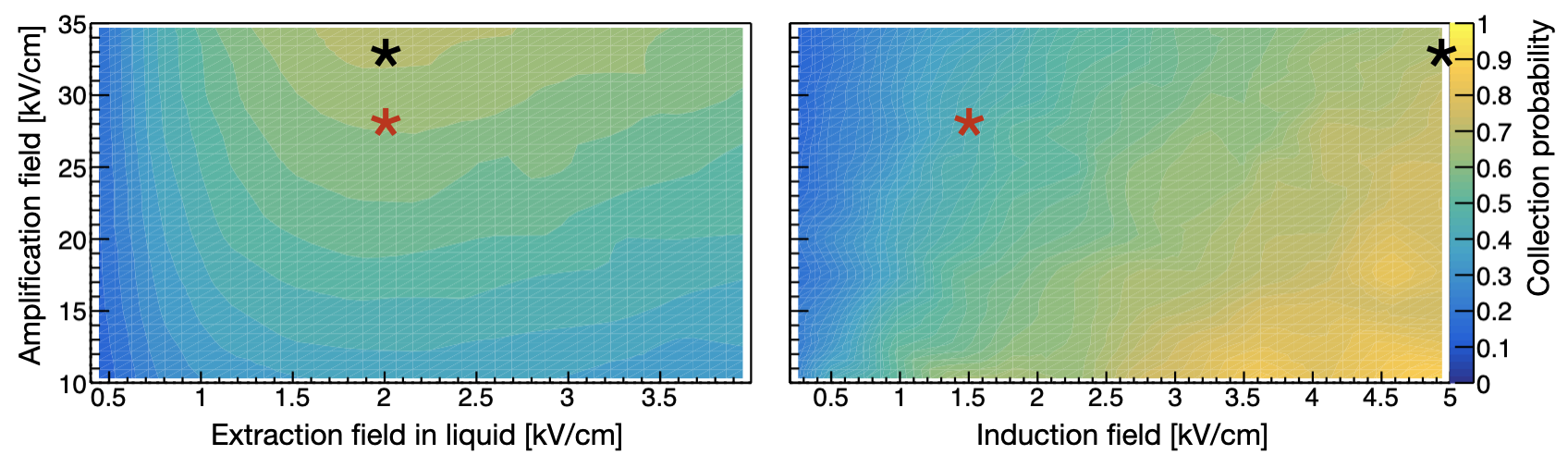}
\caption{Simulated values of $\varepsilon_{trans}$ (left) and $\varepsilon_{ind}$ (right). The red dots correspond to the best operating fields the \pilot was operated for the data taking run. The black dots indicate the field values at which the previous dual phase TPC \cite{Cantini:2013yba, Cantini:2014xza, Shuoxing-thesis} was operated under stable conditions at \Geff~=~20.}
\label{fig:extr-ind-eff}
\end{center}
\end{figure}

Performing such simulations for multiple combinations of field settings allows to generate maps of $\varepsilon_{trans}$ and  $\varepsilon_{ind}$ as a function of the amplification electric field (\Eamp). The results are shown in \figref{fig:extr-ind-eff}. 
The induction efficiency increases with the induction electric field (\Eind) and tends to plateau at values near 5~kV/cm. The value of \Eind at which this plateau is reached increases with the amplification field. The optimal setting for the extraction electric field in the liquid (\Eextr) is reached at around 2~kV/cm. Above this value, the fraction of electrons extracted from the liquid and entering the holes decreases, since more electrons are lost on the bottom LEM electrode. Also shown in the figure are markers which indicate the best operating points of the \pilot and a smaller prototype equipped with a $10\times10$~cm$^2$ LEM referred to as the \textit{3L TPC} \cite{Cantini:2013yba, Cantini:2014xza, Shuoxing-thesis}. The latter was operated stably at \Geff$\approx$ 20 with electric fields of 5, 2, 33 kV/cm for \Eind, \Eextr and \Eamp, respectively. These field settings are those reported in \figref{fig:311-geometry} and referred to as the\textit{ nominal} settings. The \pilot is operated at lower field settings of 1.5, 2.0, 28 kV/cm for \Eind, \Eextr and \Eamp which, according to the simulated maps, corresponds to a transparency of 0.4$ \times$ 0.6 $ = $24\%. 
\section{Cosmic ray data and dual phase detector simulation}
\label{sec_datamc}
\subsection{Summary of the collected data}\label{sec:data_summary}
The TPC operation ranged over a period of about 5 months. As detailed in \cite{Aimard:2018yxp}, technical issues on the absolute high voltage of the extraction grid and limitations on the maximum applied voltage of the $50\times50$~cm$^2$ LEMs prevented us from operating the TPC for a long duration and at the nominal field settings. Therefore most of the run period was dedicated to detector optimisation, and in this paper, we discuss the data from approximately 300k triggered events. Results are shown for data collected at electric field settings which optimise the $S/N$ (see \tabref{tab:data_MC_settings} for details), taking into account the above mentioned limitations. For the rest of this paper, the long (about 5 hours) stable run used for detector performance studies under stable conditions is referred to as \textit{Reference Run}. 

In \tabref{tab:data_MC_settings}, we show the settings and some properties of the analysed data-sets along with that of the detector Monte Carlo (MC) simulation tuned to match the effective gain and electron lifetime of \textit{Reference Run} as well as the detector acceptance. 

\begin{table}[h!]
\renewcommand{\arraystretch}{1.1}
\begin{center}
\begin{tabular}{>{\RaggedRight}p{.25\textwidth}p{.1\textwidth}p{.1\textwidth}p{.001\textwidth}p{.1\textwidth}p{.1\textwidth}}\\
  \toprule
  & \multicolumn{2}{c}{\textbf{Reference Run}} & &\multicolumn{2}{c}{\textbf{Field scans}}\\
 & data & MC & &Extr. &Amp.\\
  \midrule
 Ind. field [kV/cm] & 1.5 & - & &1.0 & 1.0 \\
 Amp. field [kV/cm] & 28.0\tablefootnote{The four corner LEMs are operated at 24 kV/cm \cite{Aimard:2018yxp}.} & - & &28.0 & 24-27\\
 Extr. field [kV/cm] & 2.0 & - & &1.2-2.2 & 1.5 \\
Effective gain \Geff& 1.9 & 2.0 & &0.9-1.5  & 0.5-1.2 \\
Electron lifetime [ms] &7.0& 10& & - & -  \\
$S/N_{MIP}$&12.0 & 13 & & 5.7 -- 9.5 & 3.2 -- 7.6 \\
N events ($\times10^3$)&40& 16& & 22& 213 \\
N sel. tracks ($\times10^3$) &15& 5.5 & &1.3&7 \\
    \bottomrule
       \end{tabular}
     \end{center}
     \caption{\label{tab:data_MC_settings} Summary of data taking conditions for the data-sets used in this paper. The $S/N$ is given before the application of the coherent noise filter (see text for definitions).}
   \end{table}   
   
The values of \Geff in the table are calculated from the data as explained in \secref{sec:geff} and the $S/N$ ratio is computed according to the definition of Equation \eqref{eq:s/n} for a MIP crossing the detector at ($\theta,\phi$)=($90^\circ,45^\circ$) at the quoted purity level.  

\begin{figure}[h!]
\centering
 \includegraphics[width=\textwidth]{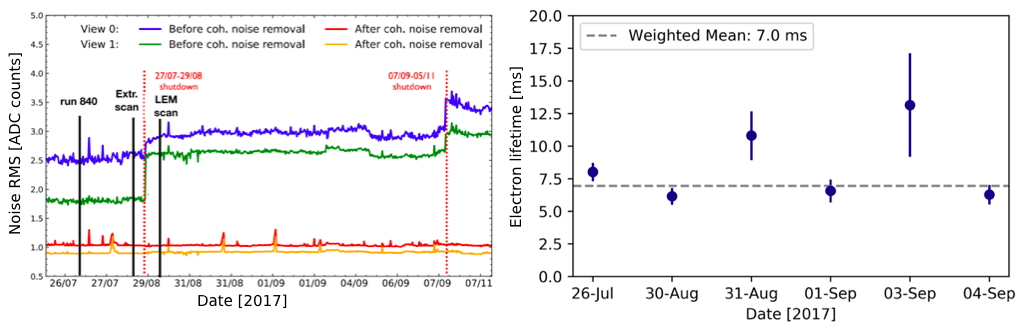}
 \caption{Left: evolution of the noise during the entire TPC operation period. The four colours correspond to the noise on the two views, before and after coherent noise removal. Right: evolution of the electron lifetime over the same data-taking period.Uncertainties are statistical only, and the large error bars on the plot are related to the different number of tracks selected for the lifetime evaluation. The weighted average of the lifetime is $\lambda = (7.0 \pm 0.4) $ ms.}
 \label{fig:noise-rms} 
\end{figure}

The evolution of the RMS of the noise on the charge readout during the operation  period of the TPC is shown on the left side of \figref{fig:noise-rms}. 
The periods corresponding to the acquisition of the data discussed in this paper are also indicated. The dashed bars indicate the two shutdown periods of approximately 1 month when maintenance was performed on the LEM high-voltage power supplies. As can be seen, the noise RMS level is significantly reduced (by more than a factor of two) by the coherent noise filter to reach a stable value of around 1.1 ADC counts (1000 electrons) in both views over the entire operation period of the chamber. The origin of the increase of the coherent noise after both shutdown periods is not well understood. It is most likely due to ground loops created during the maintenance periods on the CAEN HV supply and re-connection of sets of slow control cables. 

Detailed Fast Fourier analysis of the noise frequency discussed in \cite{andrea-thesis}, shows a prominent peak at around 900 kHz. A good stability is nevertheless observed for each period.
The right side of Figure \ref{fig:noise-rms} shows the electron lifetime measured at different dates. The measured purity was found to be stable throughout the whole data-taking period. Further information how the value given in \figref{fig:noise-rms} is computed is given in section \secref{sec:geff}.  

All the data are acquired with a trigger requiring a coincidence of all 5 PMTs. Their thresholds are set to maintain a data acquisition rate at the level of 3 Hz \cite{Aimard:2018yxp, aimard:2020qqa}. At those settings, our event sample predominantly consists of electromagnetically or hadronically induced cascades (showers), emitting large amounts of scintillation light, or MIPs, which cross most of  the fiducial volume at a relatively shallow angle. 

\subsection{Detector simulation}
\begin{figure}[h!]
\begin{center}
\includegraphics[width=0.9\textwidth, height=0.7\textwidth]{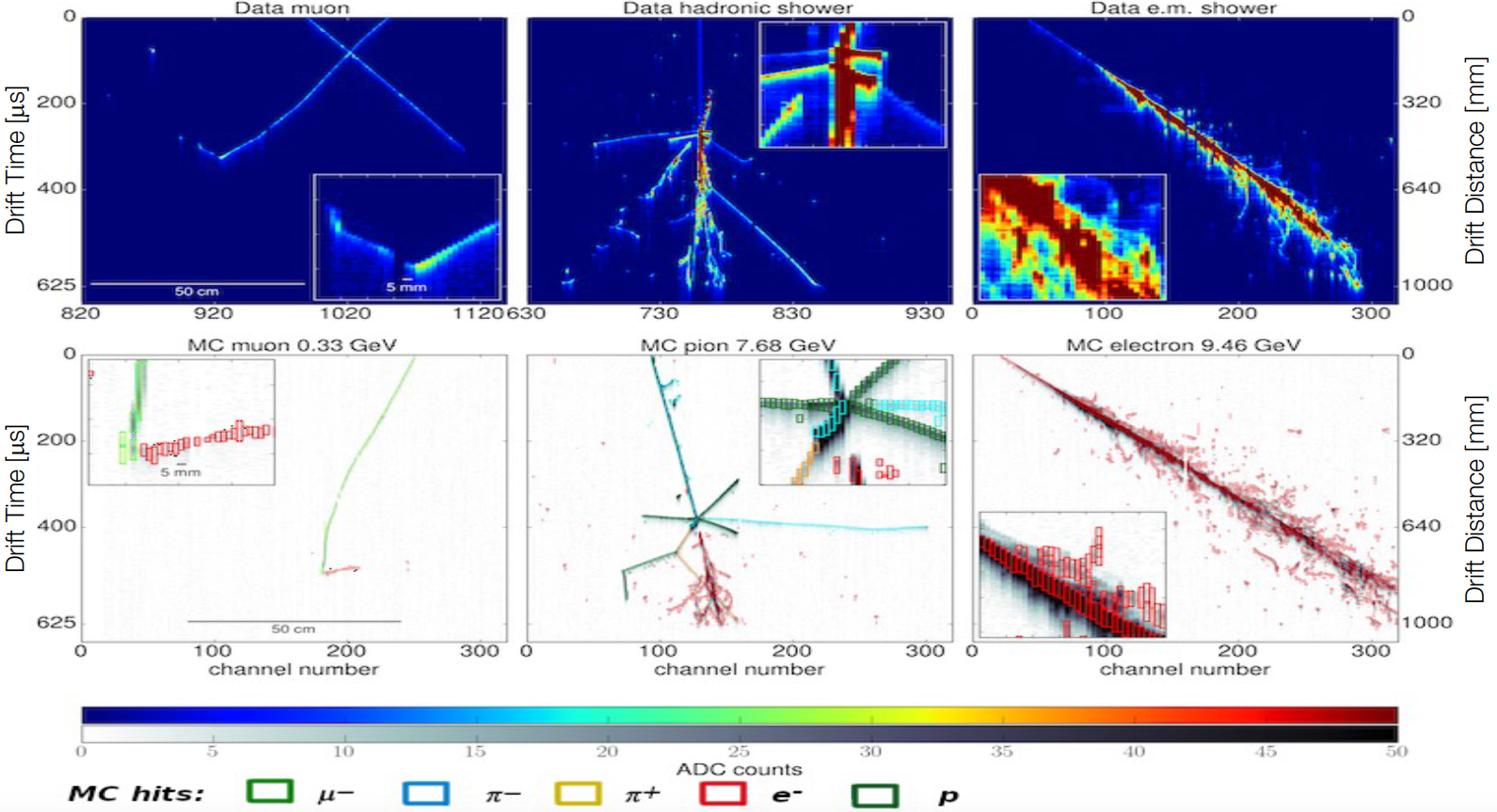}
\caption{Event displays for the three categories of events at the \textit{Reference Run} settings. The images at the top correspond to the data, those at the bottom to the Monte Carlo simulation. From left to right: a stopping muon, a hadronic and an electromagnetic shower.}
\label{fig:ExampleEvents}
\end{center}
\end{figure}

A detailed Monte Carlo simulation of the detector was developed to reproduce the detector acceptance and track topologies. Typical events for both data and MC at \textit{Reference Run} settings are shown in \figref{fig:ExampleEvents}. These include a muon decaying in the active volume where the Michel electron is clearly visible, a hadronic and an electromagnetic shower. The data are shown with a colour scale representing the hit amplitude while reconstructed MC hits are coloured  according to the particle species.


The simulation is used to check the energy range of the cosmic muon sample, verify the muon selection purity and quantify the track reconstruction efficiency, using a method explained in \secref{sec:trackrecoeff}.

Since comparison with data requires a reproduction of the atmospheric flux and simulation of the detector acceptance, an event sample  simulating the atmospheric flux was generated with the CORSIKA \cite{corsika} package. Primary particles from CORSIKA were uniformly generated on a plane, located 3 metres above the top of the TPC fiducial volume to simulate events produced from the interactions of these particles with the cryostat or detector materials.    

The trigger was simulated to reproduce the event rate, event topologies and the relative amount of tracks and showers observed in our data sample. The detector was divided into cubic areas with a volume of $25\times25\times25\,\mathrm{cm}^3$ each. From the centre of each volume, a light map was built simulating 100 million photons to compute the probability of and the arrival time for a photon produced at the centre of that volume to reach each  photomultiplier. The probability and arrival time for each position in the detector was obtained by a linear extrapolation from the simulated values at the centre of each volume, as described \cite{aimard:2020qqa}. A simplified trigger simulation was applied, which required all five PMTs to measure at least 1750 photons in a time window of 80 ns.

\begin{figure}[h!]
\begin{center}
\includegraphics[width=\textwidth]{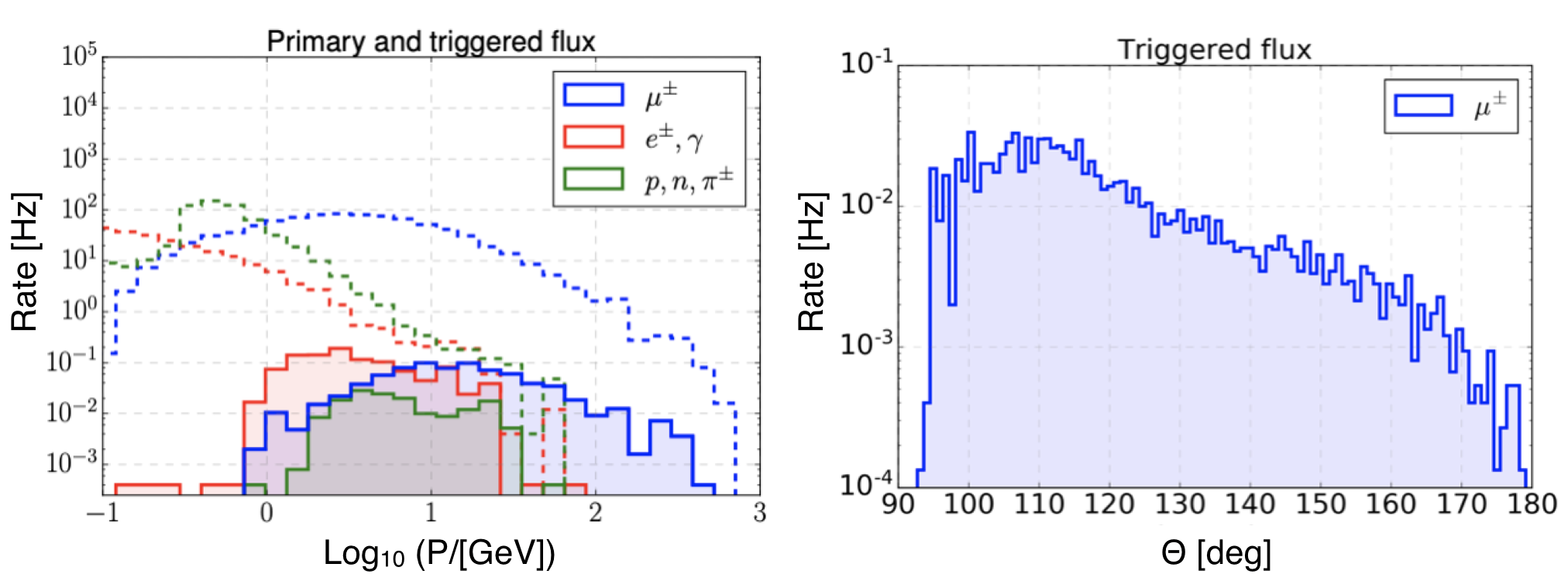}
\caption{Left: simulated rate of particles entering the fiducial volume of the TPC, in terms of expected number of particles per unit time, as a function of the log of the particle's momentum. Empty histograms with dashed lines represent the total flux, while histograms with full lines are the distribution of particles passing the trigger. Right: polar angle distribution of the muons which pass the trigger condition.}
\label{fig:corsika-tpc}
\end{center}
\end{figure}

\figref{fig:corsika-tpc} shows the momentum distribution of the generated particles  entering the TPC and those which pass the simulated trigger condition. The polar angle distribution of the triggered muons is also shown.
The particles are grouped according to the topology of events they produced when interacting in the $3\,\mathrm{m}^3$ of liquid argon target. Given the $15\,\mathrm{cm}$ radiation and $84\,\mathrm{cm}$ interaction lengths of LAr, both electromagnetic (EM) and hadronic showers initiated by electrons/gammas and hadrons, respectively, will generally be contained in the detector volume. As seen in the figure, the triggered event sample consists of muons with a broad momentum distribution peaking at around 10 GeV and with polar angles of $\sim$120 degrees, and a quite substantial content of showering events in the 1-10 GeV range. Triggered muons in those energy ranges are MIPs and after careful track selection can be used to characterise the TPC effective gain. 

\subsubsection{Waveform generation and noise simulation}
The energy deposited in the liquid argon is converted into ionisation charge. The fraction of charge lost due to electron-ion recombination is estimated with Birk's Law and the drift of the charge can be performed following an electric field map which is extracted from electrostatic simulations of the detector \cite{comsol}. Charge attenuation due to electron attachment as well as the spatial spread of the electron cloud from transverse and longitudinal diffusion along the drift are also included in the simulation according to the values shown in \tabref{tab:lar-prop}.

Once the electron cloud has reached the anode, the waveforms on each view are generated by convoluting the corresponding current with the charge readout detector response function described in \secref{subsec:readout_resp}. The effective gain is taken into account by multiplying the number of electrons inside the cloud by the desired number. In addition, the broadening of the signal width and reduction of its amplitude at low extraction fields is included in the simulation by smearing the arrival time of the electrons in the slow component according to an exponential distribution, see \figref{fig:extr-slow-component}. The time constant of the exponential is chosen so that the width of the generated waveform matches those from the extraction field scan data. 

\begin{figure}[h!]
\centering
 \includegraphics[width=0.9\textwidth,height=0.3\textwidth]{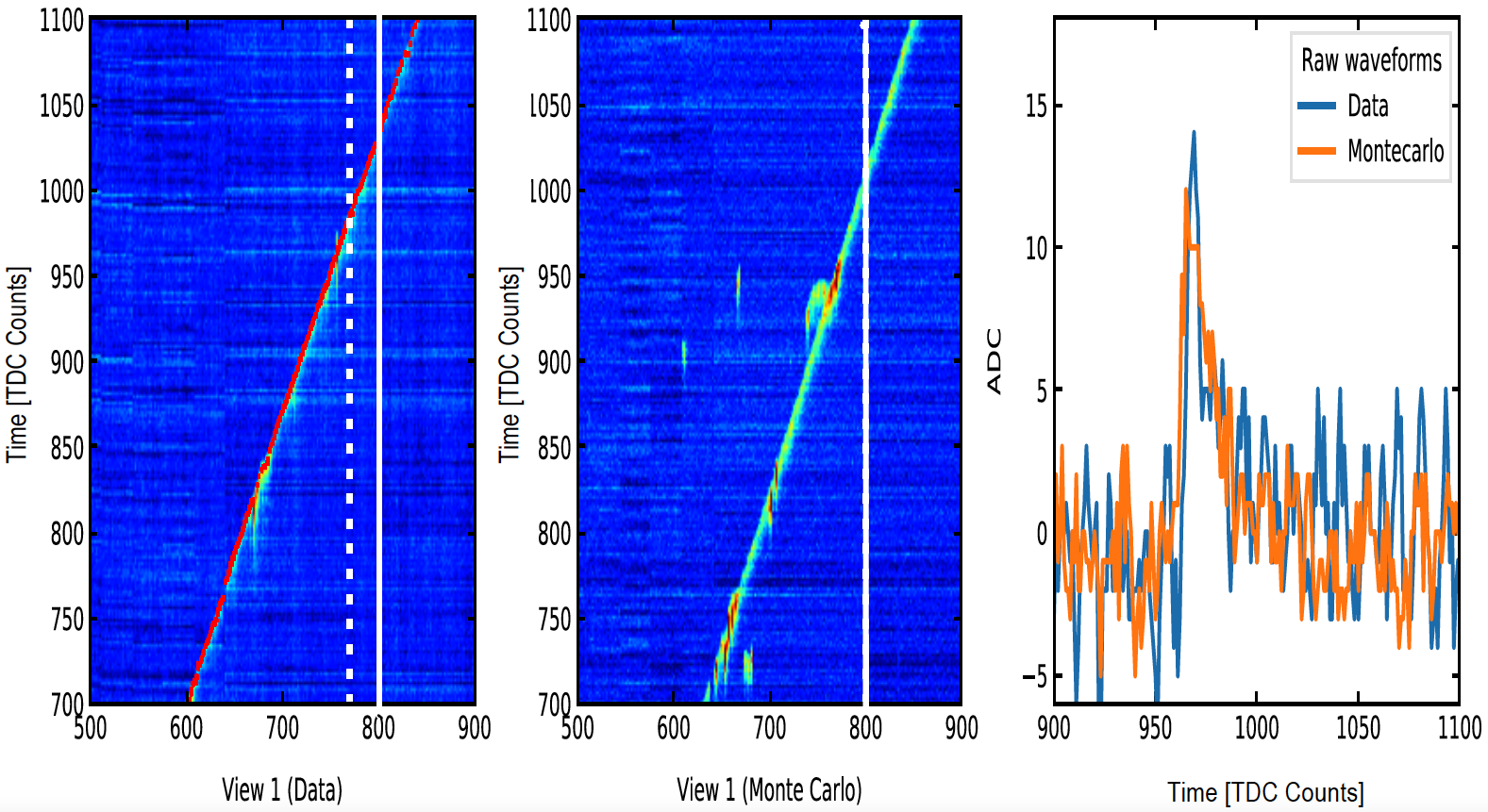}
 \caption{Event display showing a MIP candidate from data (left) and a simulated muon in the data driven simulation (centre), in the time versus channel number plane. The continuous line is due to a dead channel, while the dashed line indicates the channel from view 1 of the waveform shown in the right plot.}
 \label{fig:track_data_mc_noise} 
\end{figure}

A simulation of the coherent noise is also included. This is done by using an Inverse Fourier Transformation to produce a waveform based on the frequency spectra of the noise in data. An identical random phase is assigned to groups of channels, following the same correlation scheme that is observed in the data. Details on the waveform simulation at low extraction fields can be found in \cite{christoph-thesis}, and on the simulation of the noise in \cite{andrea-thesis}. In \figref{fig:track_data_mc_noise} we show a simulated muon event (view 1) and a MIP candidate from data at the settings of the \textit{Reference Run}. The noise pattern, the dead channels and the waveform are well reproduced in the MC.
The detector simulation features a data-driven approach to describe the waveform shape and the noise. This allows a simulation of the TPC response at any field settings and an effective gain with good agreement to data.  A comparison of the reconstructed track level between MC and data for the specific case of the \textit{Reference Run} is provided in \secref{sec_detperf}.

\subsection{Cosmic track reconstruction}\label{sec:track_reco}
\begin{figure}[h!]
\centering
 \includegraphics[width=\textwidth]{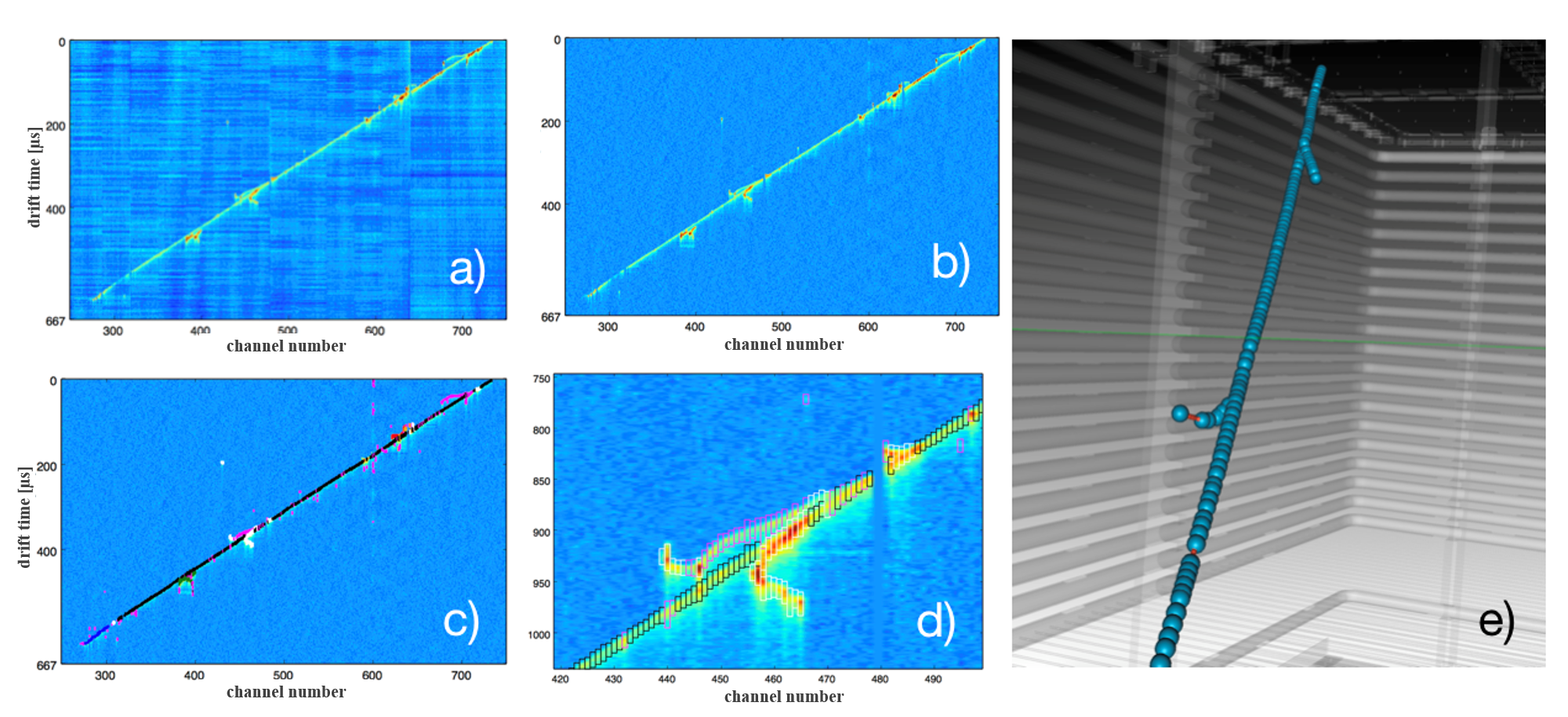}
 \caption{Schematic example of the different reconstruction stages applied to a cosmic muon track collected with the \pilot. The raw data (a) is first preprocessed in order to remove the noise (b) and to identify signal hits (c). Hits are grouped into 2D clusters (d) and clusters are finally associated to form a 3D image of the event (e).}
 \label{fig:track-3d-reco} 
\end{figure}

The off-line reconstruction of tracks from both simulation and acquired data is performed within the LArSoft software package \cite{larsoft} with the same procedure applied for data and MC events. 
The entire reconstruction procedure is illustrated  in \figref{fig:track-3d-reco} for a muon track candidate.

The first step of the event reconstruction procedure consists in identifying so called \textit{regions of interest} (ROIs) where the signal exceeds a certain threshold above the baseline pedestal. This is performed by standard threshold discrimination: the waveform is scanned for peaks above a certain average pedestal. This initial pedestal is computed in order to optimise signal over background discrimination, and the following iterative procedure guarantees that its initial value has a very small effect on the final S/N ratio.  To avoid any bias due to physical signals, the ROIs are neglected during the subsequent pedestal subtraction and noise filtering process. The absolute value and slow fluctuations of the pedestal are subtracted by applying a polynomial fit on the waveform. Noise with a periodic behaviour is removed by the successive application of Fast Fourier transform algorithms.  A coherent noise filtering algorithm is also applied to reduce the amplitude of noise patterns which are common to groups of channels at a given time. The entire process is repeated in an iterative manner since at each step the image of the event is improved and new ROIs may be identified.

At the end of this process the baseline has less fluctuations and physical hits can be extracted from the waveforms by means of standard threshold discrimination. Once hits have been identified on both views, neighbouring hits are merged together to form 2-dimensional clusters. The clusters from each view are then matched together, allowing to build \textit{3D track} objects by assigning a trajectory to the corresponding group of hits. On some occasions, clusters may be interrupted because of a malfunctioning or dead readout channel.
During the operation of the TPC, about 14 channels ($\sim$1\% of the total) were found to be problematic. In addition, each strip immediately above the 2 mm of copper guard ring surrounding the LEM borders \cite{Aimard:2018yxp} does not register a signal (four strips on view 0 and twelve on view 1). Those channels are removed from the track reconstruction algorithm to ensure the missing hits do not interrupt the track fitting procedure. 

While this approach permits reconstructing straight tracks with high efficiency, it is not designed to provide an accurate 3D representation of showering events which most of the time will be reconstructed as multiple small clusters. 
Since we have no a priori knowledge of the event topology, the procedure described above is applied on all data and shower-like events are separated from muon-track candidates by off-line tools (see \secref{sec:cut-method}). The same procedure is applied to the MC sample where the noise and detector response are simulated according to data raw events. 

\subsection{Track reconstruction efficiency\label{sec:trackrecoeff}}
\begin{figure}[h!]
\begin{center}
\includegraphics[width=\textwidth]{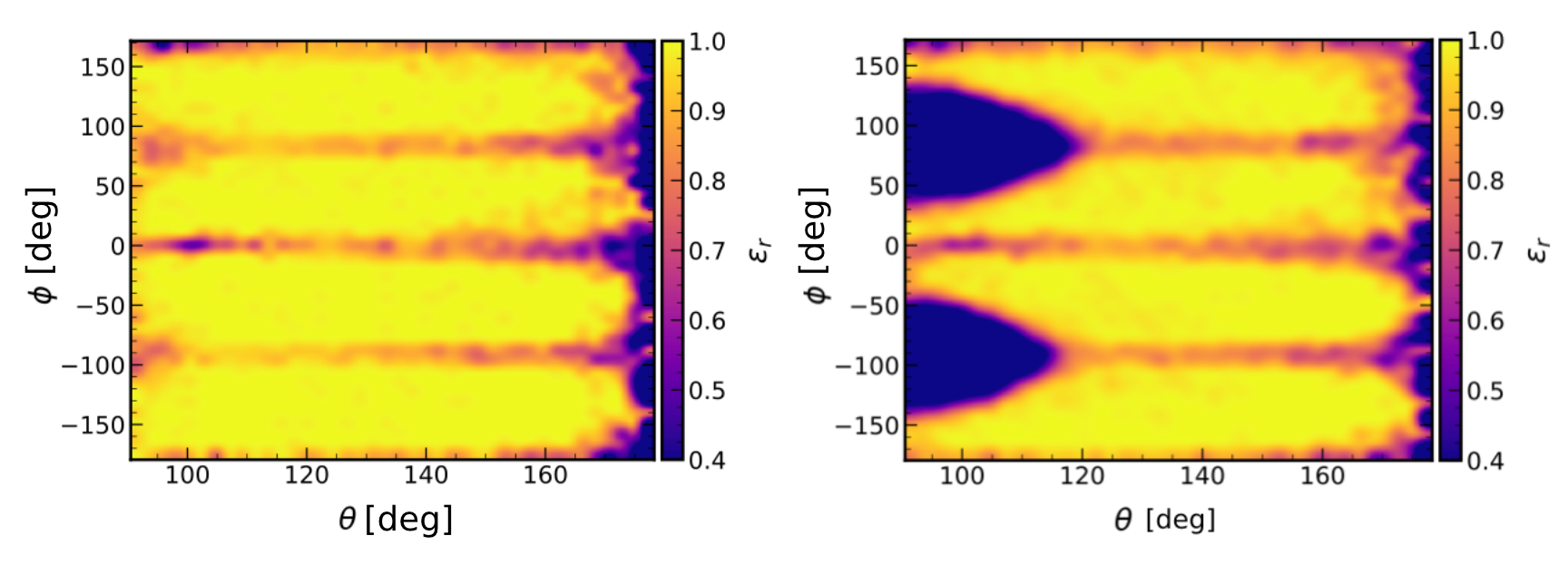}
\caption{Efficiency to reconstruct a track in 3D as a function of the polar and azimuthal angles, estimated using simulations. The left plot corresponds to the \pilot TPC operated at $\Geff$ of 2 and the right plot to the TPC operated at \Geff of 1.5. }
\label{fig:rec_eff}
\end{center}
\end{figure}

The efficiency to reconstruct tracks in 3D, $\varepsilon_r$, is estimated with the use of the flat phase space MC where  primary particles, referred to as \textit{MC-Particles}, are generated over the 4$\pi$ phase space. The noise and electron lifetimes are set to the \textit{Reference Run} settings shown in \tabref{tab:data_MC_settings}. 
The quality of the reconstruction in the simulation is computed by associating the reconstructed hits inside a track with the original energy deposit from the related GEANT trajectory. This quantity is called completeness. Tracks with completeness greater than 50\% are considered well reconstructed.
More details on the method to compute the reconstruction efficiency are provided in \cite{andrea-thesis}. The efficiency maps as a function of $\theta$ and $\varphi$ for a TPC operated at \Geff of 2 and 1.5 are shown in \figref{fig:rec_eff}. The former corresponds to the \textit{Reference Run} setting, and for the latter the lower effective gain is obtained by reducing the extraction field to 1.5 kV/cm.

As can be seen, $\varepsilon_r$ for a TPC operated at $\Geff=2$ is in general close to 100\%. Specific geometrical topologies of near vertical tracks ($\theta=180^{\circ}$) or, to a lesser extent, tracks parallel to the strips of one view ($\phi=180^{\circ},90^{\circ},0^{\circ},..$) are not efficiently reconstructed in 3D because the matching of the 2D tracks fails due to an insufficient number of hits on one or both of the views. This is a purely geometrical effect for which the effective gain at which the TPC is operated has a minimal impact. Since the 2D hits are identified, an improvement in the 2D matching procedure could increase $\varepsilon_r$ in those regions. A 25\% reduction in the effective gain translates in the same $S/N$ reduction. As a consequence the efficiency for near horizontal tracks is greatly reduced as visible on the right map of \figref{fig:rec_eff}.

In the \pilot, even though operated at a sub-optimal effective gain of around 2, cosmic tracks are therefore reconstructed with high efficiency regardless of their topology. At \Geff=1.5, the efficiency begins to drop for certain track orientations but remains near 100\% for the average polar angle of the MIPs selected by the trigger ($\theta=120^\circ$). We note that those maps are generated for the specific case of the \pilot and clearly higher effective gains are needed for larger TPCs aiming to study neutrinos with longer drifts and possibly larger noise RMS. We also do not discuss the impact of the gain on other fundamental features such as particle identification or reconstruction of low energy neutrinos.

\section{Detector performance studies}
\label{sec_detperf}
In this section the detector performance in terms of effective gain and charge sharing from the \textit{Reference run} is discussed, taking into account the technical limitations encountered during data taking. 
Qualitative comparisons of data collected at various LEM fields with data from the \textit{3L TPC} are also discussed.
As a reminder, the four corner LEMs of the TPC could only be operated stably at the maximum electric field of \SI{24}{\kilo\volt/\cm} \cite{Aimard:2018yxp}. Unless otherwise specified, the hits belonging to channels of these LEMs are excluded from the analyses.

\subsection{Muon track selection}\label{sec:cut-method}

\begin{figure}[h!]
\begin{center}
\includegraphics[width=\textwidth]{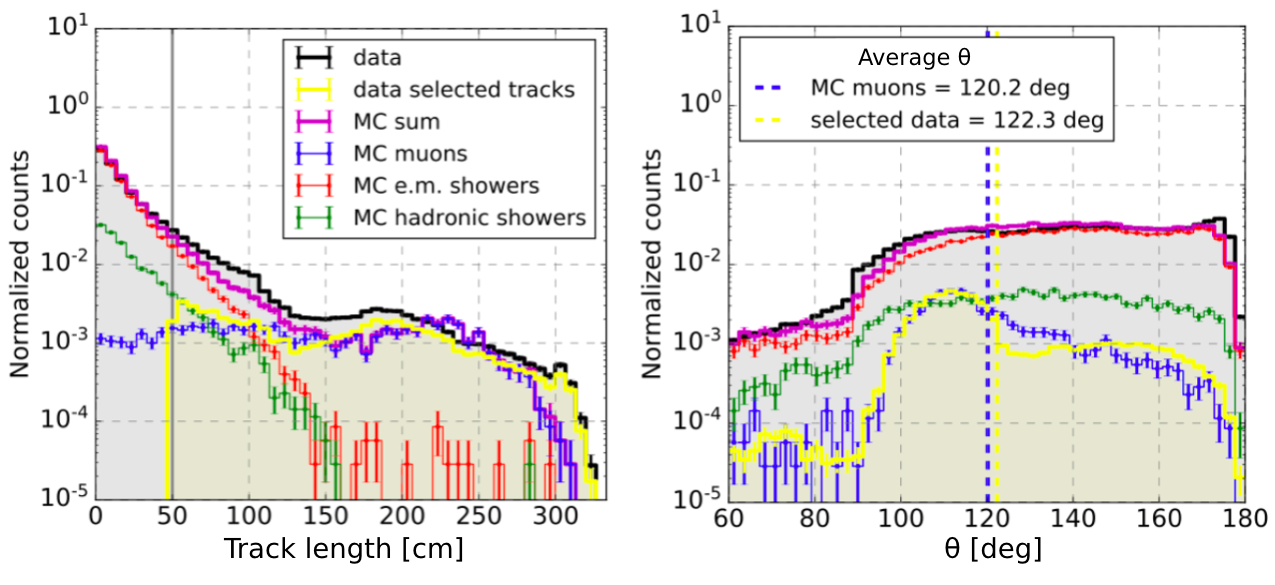}
\caption{Distribution of reconstructed track lengths and polar angles for all triggered events and selected MIPs from data compared to the Monte Carlo sample at the \textit{Reference run} settings.}
\label{fig:data_mc_l_theta}
\end{center}
\end{figure}

Through-going muons are MIPs which deposit a known $\langle dE/dx \rangle \simeq \SI{2.1}{\mega\eV/\cm}$ in liquid argon and are used to characterise the performance of the TPC in terms of effective gain. 
As explained in \secref{sec:track_reco}, showers will often be reconstructed as many independent clusters which are subsequently fitted to straight lines, producing multiple short tracks. 
The distribution of the lengths of the reconstructed tracks are shown for both MC and data in the left panel of \figref{fig:data_mc_l_theta}, where it is shown that most of the tracks associated to showers have lengths in the tens of centimetre range. The MC sample is subdivided into each particle type that belongs to the reconstructed cluster. A selection based on the length of the reconstructed track is thus an effective first step to distinguish showers from through-going muons. A cut on the track length $L\geq 50$~cm is applied, to reflect the point where an excess in data is observed over the main background due to (EM) showers.

To further clean up the sample, a second cut based on the amount of charge deposited transversely along the reconstructed 3D track is applied.
While an EM shower may occasionally be reconstructed as a single track, since its Moliere radius in LAr is about \SI{9}{\cm} \cite{pdg}, the transverse spread of the charge will be relatively large compared to a MIP track. 
In order to take advantage of this property of an EM shower, a new variable called the Charge Box Ratio ($CBR$) is defined using the fractional difference between the charge deposit in two boxes around the reconstructed track.

\begin{figure}[h!]
\begin{center}
\includegraphics[width=\textwidth,height=0.3\textwidth]{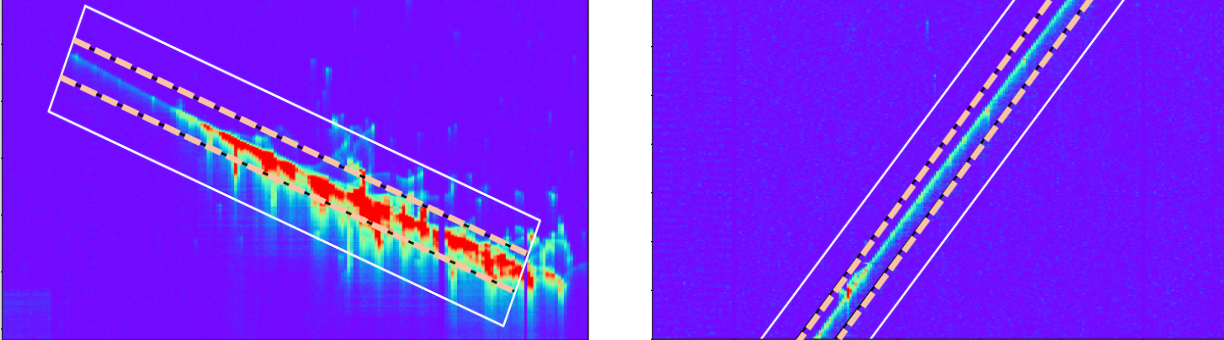}
\caption{Illustration of the algorithm which computes the charge box ratio on an EM shower reconstructed as a single track (left) and a muon candidate (right). The CBR is equal to 56$\%$ for the EM shower and $13\%$ for the muon candidate. These event displays are from the \textit{Reference run} and only view 0 is shown. The inner dashed box has a transverse dimension of \SI{3.5}{\cm} and the outer one \SI{9}{\cm}.}
\label{fig:highway_event}
\end{center}
\end{figure}

The amount of charge contained in the large white box in Figure~\ref{fig:highway_event}, $Q_{out}^i$, and the charge in the orange dashed box immediate adjacent to the reconstructed track, $Q_{in}^i$, are computed for each view, $i$.
The $CBR^i$ is then defined as the fractional difference between these two quantities with respect to the amount in the adjacent box, $CBR^i=(Q_{out}^i-Q_{in}^i)/Q_{in}^i$.
A cut of less than 20\% on the $CBR$ in both views is required to tag the cluster as a MIP track. This value is based on detailed Monte Carlo simulation studies and verified through visual scans of the event displays of candidate tracks to maximise the track selection efficiency and the muon sample purity. 

\tabref{tab:cut_track_reduction} shows the effect of the  cuts on the track lengths and $CBR$ on a MC sample that reproduces the fraction of event categories expected in data.
Of the total number of reconstructed tracks in the sample, only about 5.6\% of them are from muons before the selection criteria are applied.  
When the track length cut of $L \geq 50$~cm is applied on the whole data set, about 93\% of the EM showers and 85\% of hadronic showers are eliminated, while preserving over 93\% of the MIP-like tracks.  

When the cut on $CBR\leq20\%$ is applied, only 0.4\% and 1.9\% of the tracks from EM and hadronic showers are left, respectively.
These correspond to only 6\% and 5.5\% of the remaining tracks resulting from EM and hadronic showers respectively, and about 88.5\% from MIP.
This means that given the initial sample composition shown in the table, the muon purity increases from 5.5\% of all tracks to 88.5\% after the two selection cuts are applied. 
This also corresponds to an overall selection efficiency of 61\% for muons and over 99\% and 98\% rejection efficiency to EM and hadronic showers, respectively, clearly demonstrating the effectiveness of the selection criteria.  As potential improvement for future analyses, a more sophisticated alternative selection method based on multivariate classifiers has also been developed to take advantage of correlations between track and shower variables. 

\begin{table}[h!]
\renewcommand{\arraystretch}{1.2}
\begin{center}
\begin{tabular}{crrrr}
  \toprule
 & EM shower & HAD shower & MIP & total\\
  \midrule
 N events &8595 &1316 & 6750  & 16666 \\
 \midrule
  N reco tracks &83269  &14222  &7029 & 126092 \\
  $L\geq50$cm &5481 &2079  &6549 & 14109 \\
 $CBR\leq20\%$  &  293  &265  &4299  & 4857 \\ \midrule
Residual Fraction &  0.4 \% & 1.9 \%&  61.2\% & 3.9\%\\
        \bottomrule
       \end{tabular}
     \end{center}
     \caption{\label{tab:cut_track_reduction} The numbers of events and tracks corresponding to each category of interaction in the Monte Carlo samples. Track refers to that associated to a particle of the specified category. The residual fractions represent the percentage of tracks left in each category after the two selection cuts are applied.}
   \end{table}

   \par
 
The distribution of the selected MIP track candidates from data with the cut-based method described above is also shown in yellow in \figref{fig:data_mc_l_theta}. As can be seen, the distributions of all MC and data reconstructed tracks are in relatively good agreement before and after the selection criteria are applied, illustrating that the detector acceptance is well reproduced by the simulated trigger. The distribution of selected MIP candidates in data is close to that of the reconstructed muons in the MC.





%
%
\subsection{Space Charge and Field Distortion}

A non-uniform electric field applied on the active LAr volume of the detector worsens the energy and position resolutions and makes straight tracks appear bent and distorted.  The drifting ionisation electrons would not travel upwards along a straight line with a uniform speed and could gain a horizontal drift component in a non-uniform field.  In a LAr TPC, a non-uniform field could be caused by positive argon ions produced in the ionisation by a traversing charged track. Due to their large mass, the ionised argon atoms drift towards the cathode with much lower speed than the ionisation electrons and remain in the liquid for a much longer time. These positive ions can accumulate within the detector active volume and create a non-uniform high charge distribution that could locally distort the drift electric field. 

In addition, in a DP LAr TPC, for every electron created in the LEM avalanche, a corresponding argon ion is created. 
These ions drift towards the gas-liquid boundary, and a fraction of them could enter the liquid, adding to the space charge already residing in the active volume. The rest remains trapped on the surface of the liquid, distorting the extraction field. We define the fraction of ions crossing the liquid-gas boundary as the ion back-flow.
It is important to precisely estimate the ion back-flow, to understand how large of an impact the ions created in the gas phase would have on the space charge effect in the active liquid volume. 

\begin{figure}[h!]
\centering
\includegraphics[width=0.9\textwidth]{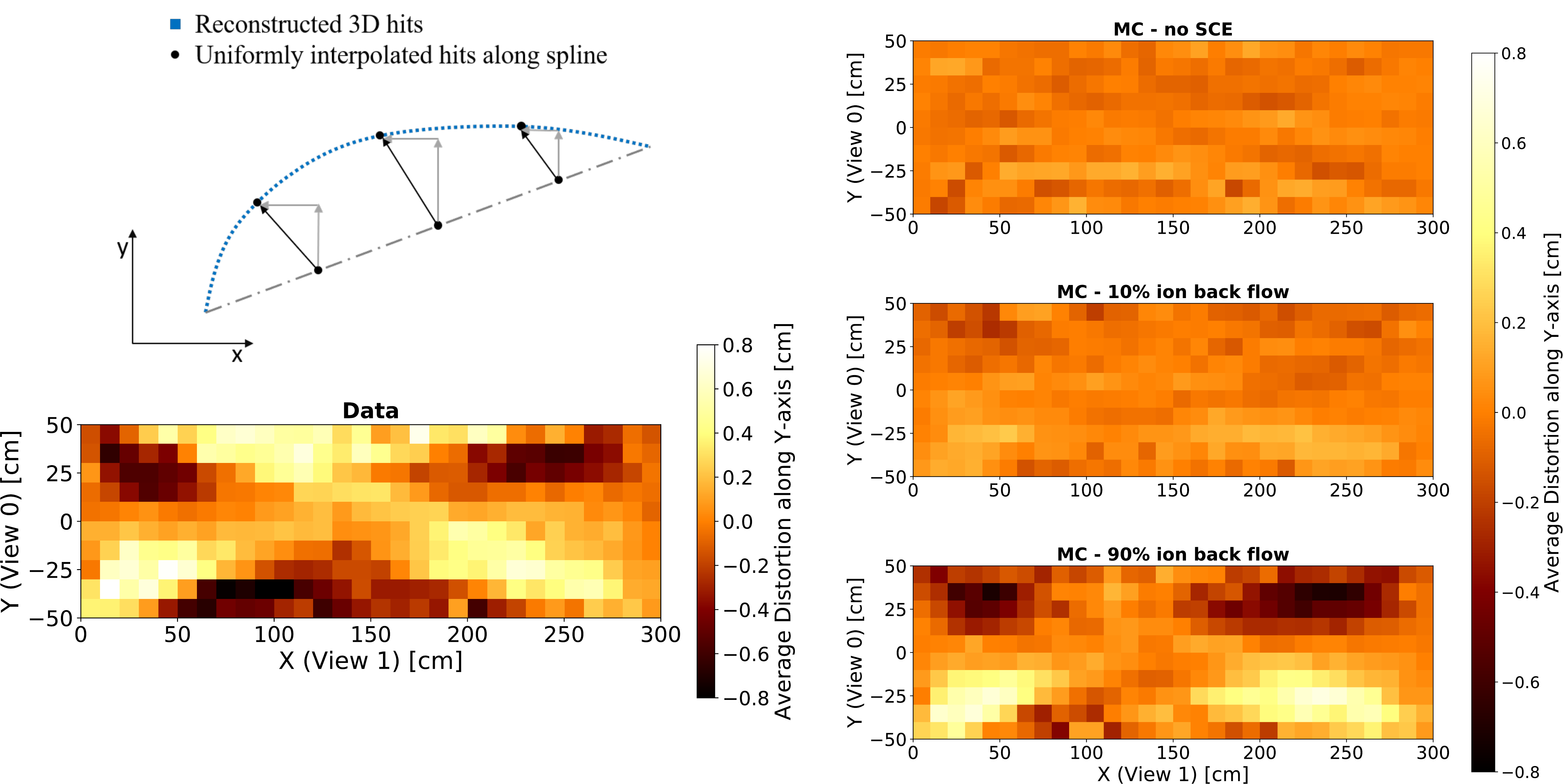}
\caption{Averaged distortions of MIP-like tracks in data (bottom left) and in simulations (right). Three hypotheses are shown for the Monte-Carlo case: no space charge (top), 10\% of ion back flow (middle) and 90\% (bottom).}
\label{fig:SpaceCharge_TopView}
\end{figure}

In order to estimate the magnitude of the space charge in the \pilot, a MIP sample was selected using the cuts described in~\secref{sec:cut-method} in the data. 
Along the reconstructed path of each track, the local 3D distortion is computed as the distance with respect to a straight line connecting the start and end points of the track, as illustrated in \figref{fig:SpaceCharge_TopView} top left.  
To determine the effect of the drift field distortions due to space charge, drift field maps with various fractions of the ion back-flow (from 0\% to 90\% in steps of 10\%) are computed using the COMSOL software \cite{comsol} with the detector technical parameters given in~\cite{Aimard:2018yxp}. The simulations consider both the space charge effect of ions coming from the gas and of ions produced in the liquid.
From each of the obtained drift field computations, the time and space displacements of drifting electrons throughout the active volume are extracted and stored in a map used as an input to the GEANT4-based simulation. 

\figref{fig:SpaceCharge_TopView} shows the average distortion of MIP-like tracks in data (bottom left), which displays pattern similar to the shape of the butterfly wings, compared to three different Monte Carlo simulations in the three plots on the right.
The top right plot shows the case in which no ions, produced either in gas or in liquid, flow towards the cathode. Hence, only the multiple scattering can affect the track path. As the resulting distortion averages at zero, it implies that the multiple scattering alone cannot explain the observation of bent tracks in the data.
The two remaining plots show 10\% (middle right) and 90\% (bottom right) ion back-flow fractions. 
As can be clearly seen, the higher the ion-back flow fraction, the better the MC samples mimic the field distortion observed in the data. 
While the 90\% ion back-flow fraction MC shows the clear emergence of the pattern similar to the data, the degree of the distortion and the detailed features, observed in the data cannot be fully described by the ion back-flow fraction alone. 
It is clear from the data plot, additional factors such as geometry and fringe effects need to be investigated to fully determine the impact of the space charge. 

Since the space charge can affect the 3D reconstruction of tracks, impacting adversely the imaging capability of the detector, this effect should be further investigated, especially for large scale future detectors with long drift distance and higher gains. In large volumes, ions will remain in the liquid for an even longer time. Additionally the targeted LEM gain in future detectors is an order of magnitude higher than the one achieved in the \three detector. Thus a high ion back-flow will add a significant amount of ions to the liquid. Both effects combined might therefore lead to significant distortions to the electric drift field. 
Due to lack of data statistics, however, only qualitative measurements of space charge and ion back-flow effects could be done with the \pilot. 
Therefore, in the following results, no corrections due to the drift field non-uniformity are applied.
%
\subsection{Effective gain, uniformity and charge sharing asymmetry}\label{sec:geff}
\begin{figure}[h!]
\centering
 \includegraphics[width=\textwidth]{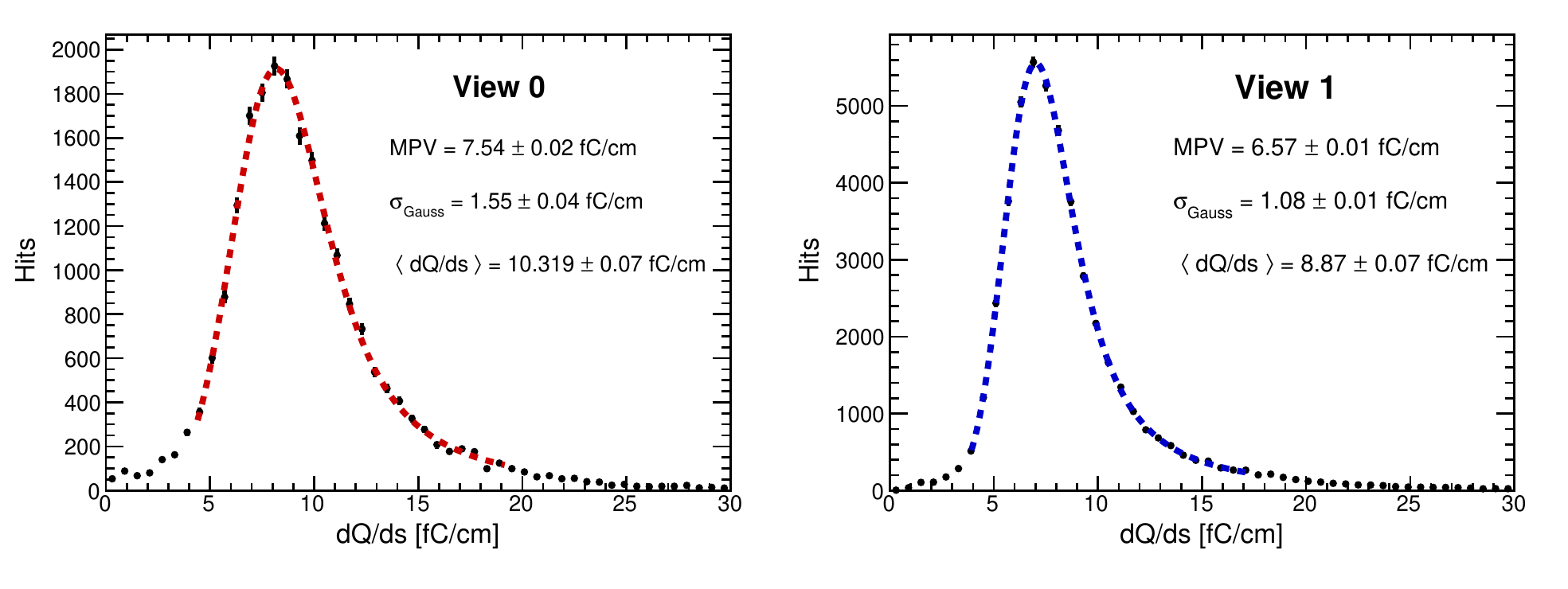}
 \caption{The $\Delta Q_0/\Delta s_0$ and $\Delta Q_1/\Delta s_1$ distributions for both views, using all LEMs together, for selected tracks from the run with electric field settings defined as the \textit{Reference run} in \tabref{tab:data_MC_settings}. The reconstructed charge in each hit is corrected for the purity effects. A Landau function convoluted with a Gaussian is fitted to the distributions and the resulting Most Probable Value (MPV) and width of the Gaussian ($\sigma_{Gauss}$) are indicated. More imperfections in the reconstruction of view 0 lead to a slightly larger width.}
 \label{fig:dqds_both_views} 
\end{figure}

The quantity $\Delta Q_i/\Delta s_i$ is the energy locally deposited by the track in liquid argon per unit length for each readout view and is used to estimate the effective gain of the chamber. 
\figref{fig:dqds_both_views} shows $\Delta Q_0/\Delta s_0$ (left panel) and $\Delta Q_1/\Delta s_1$ (right panel) for view 0 and view 1, respectively, from the sample of 3D reconstructed MIPs collected during the \textit{Reference run}.
 The effective gain is defined as the sum of the charge collected per unit length in each view divided by the average charge deposit of a MIP predicted by the Bethe-Bloch formula:
\begin{equation}
    \Geff=\frac{\langle\Delta Q_0/\Delta s_0\rangle +\langle\Delta Q_1/\Delta s_1\rangle}{\langle\Delta Q/\Delta s\rangle_\text{MIP}}
\end{equation}
Taking into account the electron-ion recombination, $\langle\Delta Q/\Delta s\rangle_\text{MIP}\sim10$~fC/cm.
The distributions shown in \figref{fig:dqds_both_views} correspond to a chamber operated at \Geff = 1.9 at the field settings and purity conditions of the \textit{Reference run} and lead to a $S/N = 12.0$ on each view for MIPs at ($\theta,\phi$) = ($90^\circ,45^\circ$). This result improves our previous measurement presented in \cite{Aimard:2018yxp}, thanks to the use of the charge readout response function described in Sec. ~\ref{subsec:readout_resp} and a more accurate muon selection.

\begin{figure}[h!]
\centering
 \includegraphics[width=\textwidth]{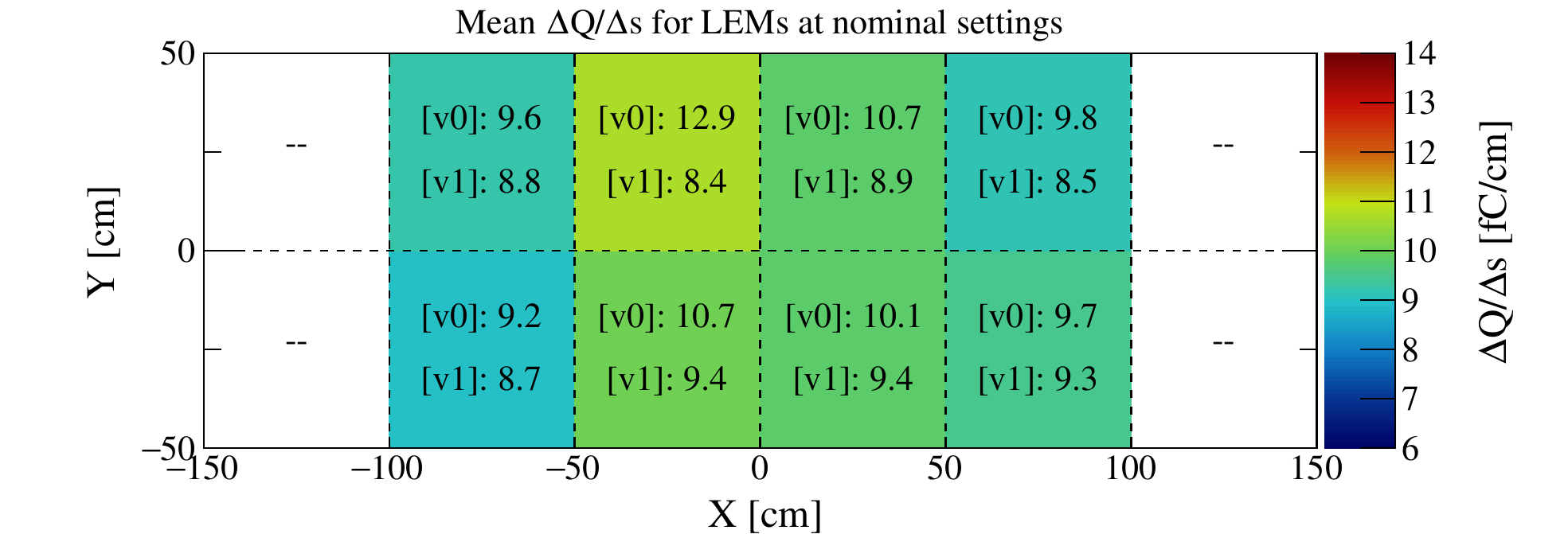}
 \caption{Average $\Delta Q/\Delta s$ collected per LEM and per view along the $x,y$ coordinates of the reconstructed tracks. The 4 corner LEMs are operated at 24 kV/cm and are excluded from the analysis while the 8 central ones at 28 kV/cm. The color scale indicates the $\Delta Q/\Delta s$ averaged over the two views.}
 \label{fig:gain-unif} 
\end{figure}

The CRP is composed of \fifty units, thus we should ensure that the effective gain is uniform over the $3\times1$~m$^2$ area. The uniformity of the effective gain is illustrated in \figref{fig:gain-unif} that shows the collected charge per unit length in both views ($\Delta Q/\Delta s$) averaged over each LEM as a function of the $x,y$ coordinates of the reconstructed MIPs. 

The fluctuation of the effective gain in the area corresponding to the 8 central LEMs is contained within 5\% tolerance.

\begin{figure}[h!]
\centering
 \includegraphics[width=\textwidth]{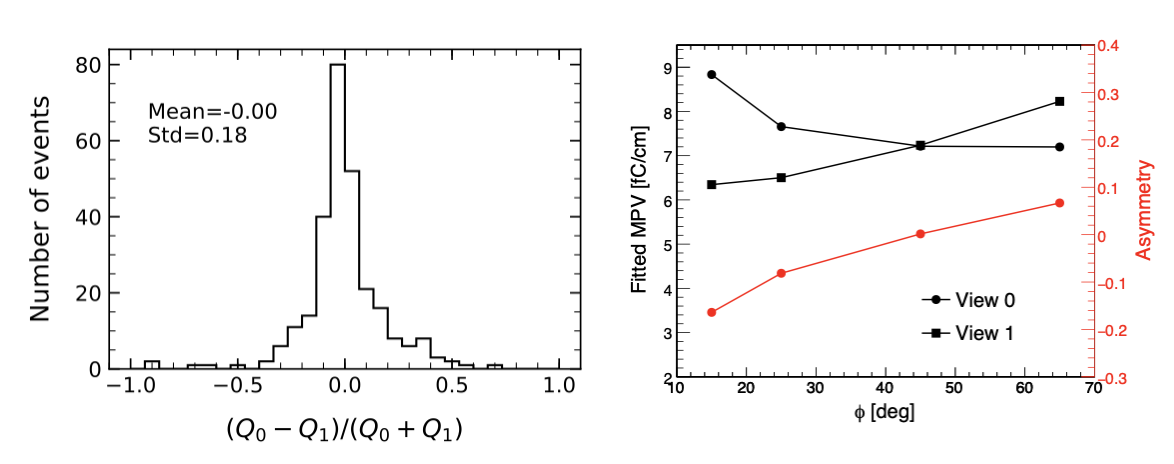}
 \caption{Left: The charge sharing asymmetry distribution for tracks in the azimuth angle range $\phi=45\pm5^{\circ}$, Right: MPVs from the fit to the charge deposit in view 0 and view 1 (in black)}
 \label{fig:dqds_asym} 
\end{figure}

In addition to demonstrating a good uniformity of the charge readout on a large area, it is important to verify the good charge sharing between the two views.
The charge sharing asymmetry coefficient is defined as:
\begin{equation}
    \mathcal{A}=\frac{\langle\Delta Q_0/\Delta s_0\rangle -\langle\Delta Q_1/\Delta s_1\rangle}{\langle\Delta Q_0/\Delta s_0\rangle +\langle\Delta Q_1/\Delta s_1\rangle}.
\end{equation}
This coefficient depends on the azimuthal angle $\phi$ and varies within $\pm1$, as shown in \figref{fig:dqds_asym} right. 
In particular, when $\phi=\pm45^{\circ}$, $\pm 135^{\circ}$, the track unit length $\Delta s$ is the same in both views. In this case, the collected charge in each view should be equivalent and $\mathcal{A}$ should therefore be equal to 0 (see \figref{fig:dqds_asym} left).

  Despite the selection operated by the trigger conditions and difficulties in the reconstruction of tracks parallel to the strips, the asymmetry remains within 15\%. 
A more careful evaluation of the systematic uncertainties on the effective gain measurement and its stability as well as on the uniformity of the CRP response will be carried out in the ProtoDUNE-DP detector, to ensure the good fitting of the LAr TPC dual phase technology within the requirements set by the DUNE experiment.

\subsection{Scans of extraction and LEM fields}

Field scans enable the study of the detector response variations as a function of the high-voltage settings and the interplay of the different fields in the final effective gain described in \secref{sec:effgain}.
We consider here the scans of the extraction and the amplification fields summarised in \tabref{tab:data_MC_settings}. The results from these scans, even if performed at sub-optimal settings and in restricted electric field ranges, can be used to qualitatively confirm the electron transmission efficiency at the liquid-gas interface shown in \figref{fig:extr-slow-component} and to quantify \GLEM as a function of the amplification field. The conditions of the field scans do not allow to study independently each of the fields. The combination of the field scans together with the simulated maps from \figref{fig:extr-ind-eff} allow us to extract the extraction efficiency and the amplification inside the LEMs in the detector.

 \begin{figure}[h!]
\centering
 \includegraphics[width=\textwidth]{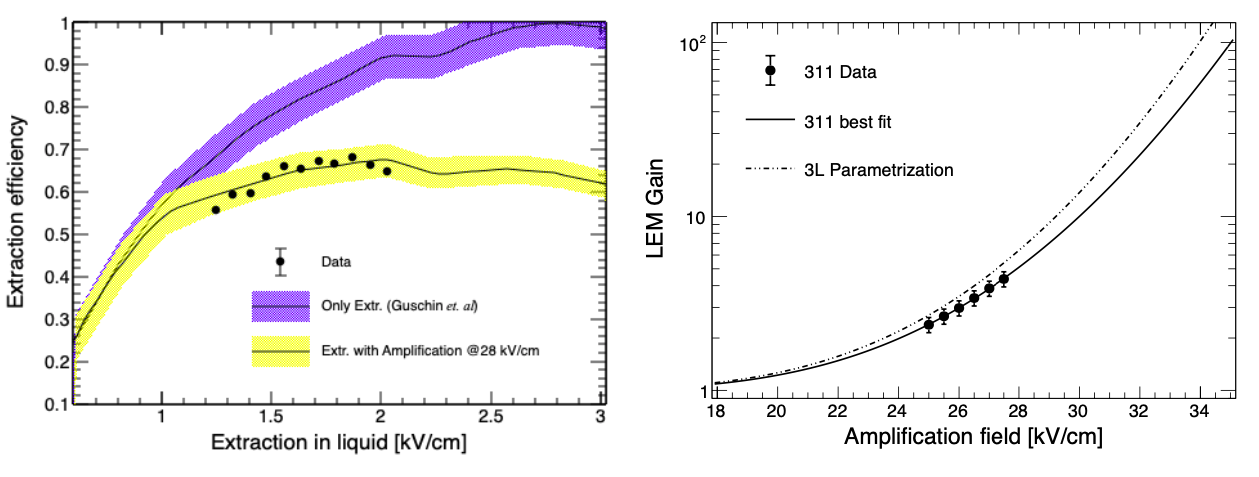}
 \caption{Left: liquid to gas transmission and extraction efficiencies as a function of the extraction field in LAr measured in the \pilot.  The measured points at 28 kV/cm obtained from the extraction scan are superimposed on the transmission prediction (yellow band), the convolution of the extraction field and the LEM transparency. Right: \GLEM as a function of the amplification field (solid circles) measured in the \pilot is compared to the best fit of the \textit{3L TPC} data (dotted line) which was equipped with a single $10\times10$~cm$^2$ LEM, both normalised to the transparency of each detector. }
 \label{fig:extr_LEM_scan} 
\end{figure}

The effective gain is defined as $\Geff=\trans\cdot$\GLEM (see Eq.~\eqref{eq:g_eff} in \secref{sec:effgain}), where the transparency is defined as in Eq. ~\eqref{eq:gain_trans} by the product between transmission and induction efficiencies. Due to the high voltage limitations, we could not operate the detector at induction fields above 1.5~kV/cm. However, we were able to study the transmission efficiency, the product between the extraction efficiency of electrons from the liquid to the gas phase and the efficiency of the electrons successfully passing through the LEM holes, $\varepsilon_{extr}^{LEM}$. The blue region of \figref{fig:extr_LEM_scan} left shows the expected evolution of the extraction efficiency from the liquid to the gas, $\varepsilon_{extr}^{liq}$, as a function of the extraction field in liquid according to \cite{Gushchin:1982}, as explained in \secref{subsec:charge_extr}. The yellow band is the product of this efficiency ($\varepsilon_{extr}^{liq}$) with the $\varepsilon_{extr}^{LEM}$ obtained from the simulated maps described in \figref{fig:extr-ind-eff} for 28~kV/cm. Data from the extraction scan shows a good agreement between the condition at which the demonstrator was operated and our prediction.

A scan of the amplification field in the limited high voltage region available was performed in order to study amplification inside the LEM holes and the resulting \GLEM. The right panel of \figref{fig:extr_LEM_scan} shows \GLEM as a function of the amplification field for the scan performed in the \pilot compared to the results from the smaller \textit{3L TPC} scan in Ref.~\cite{Shuoxing-thesis}. \GLEM is extracted by using Eq.~\eqref{eq:g_eff}. The transparency is computed from \figref{fig:extr-ind-eff} in the case of the \pilot data and the value for the \textit{3L TPC} is obtained from \cite{Shuoxing-thesis}. The two results show the same trend as a function of the amplification field.
Since the two detectors were operated at different pressures and temperature (the \pilot was operated at 1000~mbar, while the \textit{3L TPC} ran at 980~mbar), extraction and collection field, we do not expect the two fitted curves for the LEM gains to fully overlap. 

The relative gain difference between the two curves is $\approx 20\%~(40\%)$ at 28~kV/cm (33~kV/cm). 
It can be partially explained by a 2\% residual variation in density between the two detectors according to \figref{fig:gain-var}. Another contribution could come from charging up of the dielectric layer of the LEMs. While the data from the \textit{3L TPC} scan was collected before the dielectric layer of the LEM was fully charged up, we cannot conclude the same in the \pilot. The high voltage stability issues prevented operating the \pilot for a sufficiently long time to study properly the charging up effect of \fifty LEMs. It was observed in $10\times10$~cm$^2$ LEM modules that the effective gain reduces by a factor of about 3 after a characteristic time of about 1.5 days when operated at $\Geff\approx100$ \cite{Cantini:2014xza,Cantini:2016tfx, Shuoxing-thesis}.  

The results from these scans suggest that the electron extraction from the liquid to the gas is similar to previous measurements. The electron avalanche in the LEMs appears to be similar to that measured with smaller $10\times10$~cm$^2$ devices.  Additional field scans in future prototypes, such as ProtoDUNE DP LAr TPC demonstrator, would bring the opportunity for a better evaluation of the performance of the \fifty LEMs.

\section{Conclusion}
\label{sec_conclusion}
This paper presents the performance of the \pilot during its cosmic ray data taking operation in summer 2017. While the high voltage limitations prevented reaching the optimal expected amplification performance, many properties of the detector could be studied. 

 A data driven detector simulation including a realistic trigger based on the photo-detection system has been developed to reproduce the topology of the recorded tracks. The simulation has been used to tune an algorithm for track-shower separation, which is a crucial task to measure most critical performance metrics of the detector. The algorithm is based on the fractional charge difference deposited in two boxes surrounding the reconstructed track with a muon purity selection with respect to the cosmic ray sample of around $\sim$90\%. Multivariate analysis techniques can be used in the future to further improve the purity of the selected sample. A data-driven Monte Carlo simulation was used to study the impact of both signal-to-noise and geometrical effects on the efficiency of cosmic muon track reconstruction.

The study of the straight muon-like tracks allowed us to measure the drift electron lifetime which is connected to the LAr purity. We obtain a weighted average lifetime of 7 ms consistent to that in \cite{Aimard:2018yxp} and stable over time. From the selected tracks, an effective gain of 1.9 was measured. 
The differences with respect to the number quoted in \cite{Aimard:2018yxp} are due to the improved track selection and the use of a different electronic response. The charge sharing asymmetry between the two views has been studied as a function of the azimuthal angle. In the case of $\phi=45^{\circ}$, the asymmetry is centered in 0, highlighting equally charge sharing among the collection views. 

As described in \cite{Aimard:2018yxp}, we confirm the extraction of electrons from liquid to the gas phase in large areas of 3~m$^2$. Despite the lack of statistics, we address qualitatively the interplay among the different fields involved in a dual phase detector, especially between the extraction and amplification fields.

The amplification gain in the LEMs has been measured and compared to the \textit{3L TPC}. A similar trend in gain variation as a function of the amplification field is observed. However, the \three results have lower values, which could be explained by a 2\% density variation between the two setups. 


Despite the limitations, the results shown in this paper seem to confirm qualitatively the key aspects of the dual-phase technology such as the extraction of the electrons from liquid to gas and their amplification though the entire one-squared-metre {\it CRP}. Also the gain stability, purity and charge sharing between readout views looks in reasonable agreement from design expectations. However, an extensive study of the amplification process and its stability over a large readout surface during an extended period of time is still necessary to ensure the fulfilling of all the requirements for a multi-kilotonne neutrino detector. Although these
results have set another milestone towards the understanding of the dual phase technology,
future prototyping efforts are expected to confirm them under more controlled experimental
conditions and with improved reconstruction techniques. 



\acknowledgments
We are truly grateful to the strong and continuous support of CERN for the experimental infrastructure, the development of the cryogenic system and for the detector operation. We are also thankful to the CERN IT department and the IN2P3 Computing Center (CC-IN2P3) for the computing resources needed for data storage, processing, analyses and simulations. \\
This work would not have been possible without the support of the Swiss National Science Foundation, Switzerland; CEA and CNRS/IN2P3, France; KEK and the JSPS program, Japan; Ministerio de Ciencia e Innovaci\'on in Spain under grants FPA2016-77347-C2, SEV-2016-0588 and MdM-2015-0509, Comunidad de Madrid, the CERCA program of the Generalitat de Catalunya and the fellowship (LCF/BQ/DI18/11660043) from ``La Caixa'' Foundation (ID 100010434); the Programme PNCDI III, CERN-RO, under Contract 2/2020, Romania; the U.S. Department of Energy under Grant No. DE-SC0011686.
This project has received funding from the European Union's Horizon 2020 Research and Innovation program under Grant Agreement no. 654168.
The authors are also grateful to the French government operated by the National Research Agency (ANR) for the LABEX Enigmass, LABEX Lyon Institute of Origins (ANR-10-LABX-0066) of the Universit\'e de Lyon for its financial support within the program "Investissements d'Avenir" (ANR-11-IDEX-0007).

\bibliographystyle{JHEP}
\bibliography{Sections/bibfile}

\end{document}